\documentclass[superscriptaddress, amsmath,amssymb,
 aps, pra, twocolumn, longbibliography]{revtex4-2}
\usepackage{txfonts}

\usepackage{graphicx}
\usepackage{dcolumn}
\usepackage{bm}
\usepackage{xcolor}
 
\usepackage{hyperref}
\hypersetup{colorlinks=true, citecolor=blue, linkcolor=blue}
\usepackage{nicefrac}

\newcommand{\customsection}[1]{\vspace{0.3cm}\noindent\textbf{#1}}

\newcommand{\customsubsection}[1]{\noindent\emph{#1}}
\newcommand{\supop}[1]{\mathcal{#1}}
\newcommand{\eg}[1]{#1}
\newcommand{\bra}[1]{\langle #1 |}
\newcommand{\ket}[1]{| #1 \rangle}
\newcommand{\braket}[2]{\langle #1 | #2 \rangle}
\newcommand{\ketbra}[1]{| #1 \rangle\langle #1|}
\newcommand{\Tr}{\mathrm{Tr}}
\newcommand{\id}{\mathbb{I}}

\newcommand{\layer}{\ell}
\newcommand{\circuit}{c}
\newcommand{\gate}[1]{\texttt{#1}}

\renewcommand{\vec}[1]{\boldsymbol{#1}}

\begin{document}

\title{Efficient simulation of Clifford circuits with small Markovian errors}
\author{Ashe Miller}
\affiliation{Quantum Performance Laboratory, Sandia National Laboratories, Livermore, CA 94550}
\author{Corey Ostrove}
\affiliation{Quantum Performance Laboratory, Sandia National Laboratories, Albuquerque, NM 87185}
\author{Jordan Hines}
\affiliation{Quantum Performance Laboratory, Sandia National Laboratories, Livermore, CA 94550}
\author{Robin Blume-Kohout}
\affiliation{Quantum Performance Laboratory, Sandia National Laboratories, Albuquerque, NM 87185}
\author{Kevin Young}
\author{Timothy Proctor}
\email{tjproct@sandia.gov}
\affiliation{Quantum Performance Laboratory, Sandia National Laboratories, Livermore, CA 94550}
\date{September 2023}
\date{\today}

\begin{abstract}
Classical simulation of noisy quantum circuits is essential for understanding quantum computing experiments.  It enables scalable error characterization, analysis of how noise impacts quantum algorithms, and optimized implementations of quantum error correction. However, most existing efficient simulation techniques can only simulate the effects of stochastic (incoherent) noise.  The lack of efficient ways to simulate coherent errors, which are common and significant in contemporary quantum computing systems, has frustrated research.  We remedy this gap by introducing an efficient algorithm for approximate simulation of Clifford circuits with arbitrary small errors (including coherent errors) that can be described by sparse $n$-qubit Lindbladians.  We use this algorithm to study the impact of coherent errors on syndrome extract circuits for distance-3, 5, 7, 9, and 11 rotated surface codes, and on deep random 225-qubit circuits containing over a million gates.
\end{abstract}

\maketitle

\customsection{Introduction.} Classical simulation of quantum circuits lies at the heart of quantum computing theory.  Arbitrary \textit{noiseless} (unitary) $n$-qubit quantum circuits cannot be simulated by any efficient (polynomial in $n$) classical algorithm unless BQP = BPP, in which case quantum computers offer no exponential advantage.  But efficient classical algorithms can simulate key special classes of circuits \cite{Takahashi2020-yl, Gidney2021-ef, Aaronson2004-ab, Delfosse2023-ox,  Gosset2024-ez, Fontana2023-ci, Gonzalez-Garcia2024-ri, Xun2018-rp}, most notably \textit{Clifford circuits} composed entirely of Clifford gates \cite{Gidney2021-ef, Aaronson2004-ab, Delfosse2023-ox,  Gosset2024-ez}.  Clifford circuits do not offer algorithmic speedups \cite{Aaronson2004-ab}, but they can perform many useful quantum computing tasks including teleportation \cite{Bouwmeester1997-rn}, quantum error correction \cite{Google-Quantum-AI-and-Collaborators2025-ad}, randomized benchmarking \cite{Magesan2011-hc, Proctor2019-gf, Hines2024-qe}, gate tomography \cite{Blume-Kohout2017-no, Madzik2022-jh}, and the creation of entangled resource states for computation \cite{Briegel2009-fs} and metrology \cite{Bollinger1996-cb, Bao2024-bi, Proctor2018-lk}.

Adding noise to Clifford circuits---i.e., each Clifford gate causes unwanted errors---complicates simulability.  The cost of simulation depends on the type of noise.  If the noise is described by \textit{stochastic Pauli} (or Clifford) processes \cite{Hashim2024-om}, then efficient classical algorithms for \textit{weak} simulation (generating samples from the outcome distribution) of Clifford circuits exist \cite{Gidney2021-ef}, although strong simulation (exact calculation of outcome probabilities) is not known to be possible.  But stochastic Pauli or Clifford noise is a very special case.  More general noise processes are described by arbitrary completely positive trace preserving (CPTP) maps \cite{Blume-Kohout2022-ln, Hashim2024-om}. No $\mathrm{poly}(n)$ classical algorithms for simulating such noise are currently known. Full density matrix simulation requires multiplying $4^n\times 4^n$ transfer matrices, and state of the art methods \cite{Bennink2017-rr, Chen2021-eu, Pednault2017-oa} still scale exponentially. Pauli propagation techniques using quasiprobability distributions \cite{rall2019simulation, angrisani2025simulating} are efficient under some conditions, but only for incoherent classes of errors.  The simple and interesting special case of coherent (unitary) errors \cite{Hashim2024-om} is generally infeasible---Clifford gates with coherent errors become computationally universal \cite{Nebe2006-sd}, and are therefore unlikely to admit unconditionally efficient simulation. There are algorithms to simulate circuits comprising mostly Clifford gates with a small number of non-Clifford $T$ gates \cite{Bravyi2016-hq, Huang2019-ma, Kocia2020-fm, Smith2023-mu, Park2024-nr, Bu2019-xg}, but they are not suited to the situation where \emph{every} Clifford gate is slightly perturbed by coherent noise.

This is too bad, because simulating circuits with noisy gates is essential to many quantum computing tasks.  Examples include understanding quantum error correction \cite{Gidney2021-ef, Katsuda2024-hh, Bravyi2018-oi, Venn2020-of, Pato2025-qm, Venn2023-ap, Behrends2024-fm, Behrends2024-ok, Marton2023-il, Pataki2024-gx, Suzuki2017-jh}, exploring the noise-resilience of quantum algorithms \cite{Fontana2021-cy, Hann2021-zw}, fitting noise models to data \cite{Blume-Kohout2017-no, Madzik2022-jh, Malekakhlagh2025-ar}, performing noise-aware circuit compilation \cite{Cincio2021-oi}, and assessing error mitigation methods \cite{Ferracin2024-zs, Nation2021-by}. Understanding the impact of coherent errors is especially important. They are common, but very unlike stochastic errors, because they interfere constructively and/or destructively throughout a quantum circuit \cite{Proctor2021-wt, Debroy2018-an}. The absence of efficient simulation methods for qubits experiencing general coherent errors has limited explorations of these effects except for around 60 or fewer qubits \cite{Katsuda2024-hh, Pednault2017-oa, Wu2019-mb}, where full vector-state simulation is feasible; to shallow circuits with limited connectivity, where methods that do not store the entire state vector can decrease simulation costs \cite{Boixo2017-hn, Guo2019-vb, Chen2018-qc, Chen2018-rh} (e.g., noisy random circuits with 27 layers on 144 qubits have been simulated \cite{Chen2018-rh});  or to circuits containing coherent errors with special structures---such as surface codes with single-qubit phase ($z$-axis) errors, where efficient simulation is possible \cite{Bravyi2018-oi, Venn2020-of, Pato2025-qm, Venn2023-ap}, or surface codes with general single-qubit errors, where $\mathcal{O}(\exp(n^{\nicefrac{1}{2}}))$-scaling simulation is possible using tensor networks \cite{Darmawan2017-mm}.

\begin{figure*}[t!]
\centering
\includegraphics[width=18cm]{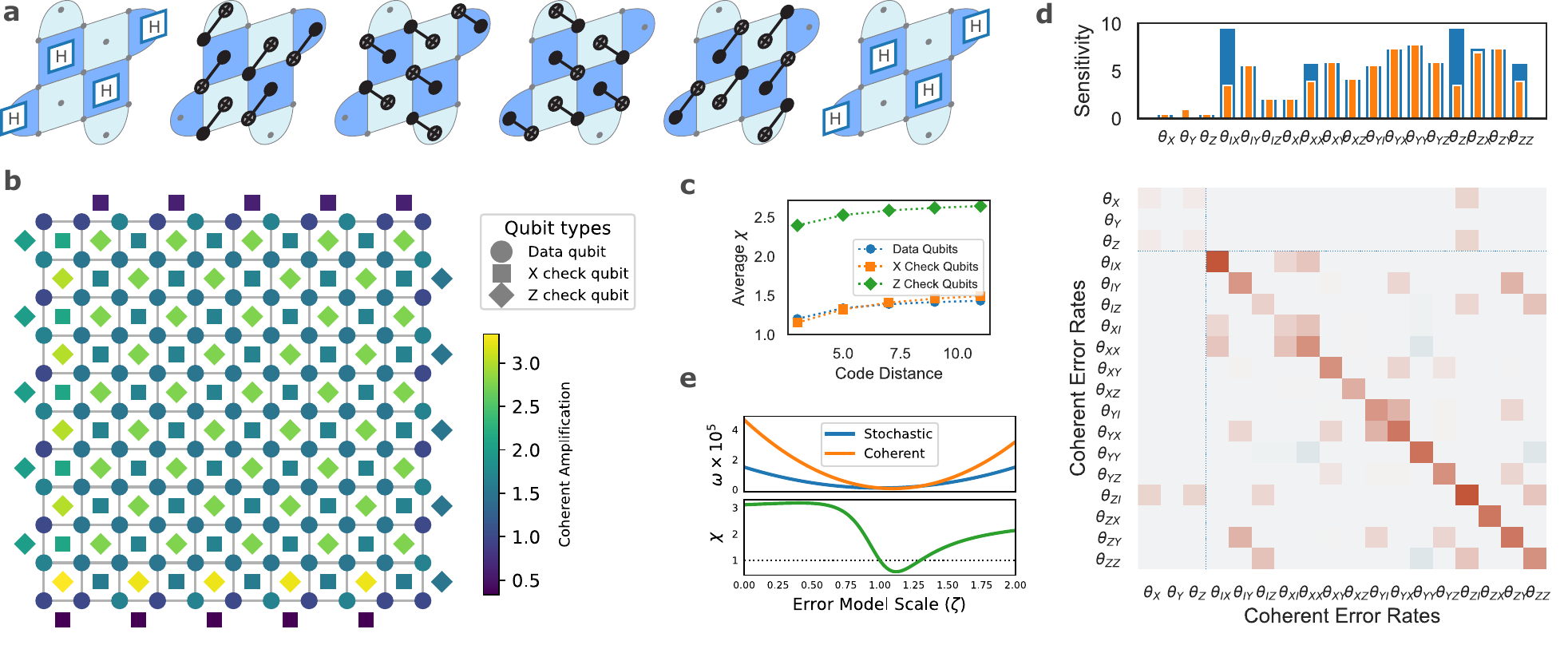}
\caption{\textbf{Coherent amplification and suppression of errors in quantum error correction circuits}. Using our efficient simulation algorithm, we studied the effect of coherent errors in syndrome extraction circuits for the distance-$3,5,7,9$ and $11$ rotated surface codes, finding that coherence can both amplify and suppress error rates by significant factors in syndrome extraction circuits. (a) shows the syndrome extraction circuit for distance $d=3$. We simulated a coherent noise model with 18 parameters $\vec{\theta}$---three describing errors on Hadamard gates ($\theta_X, \theta_Y, \theta_Z$) and 15 describing errors on CNOT gates $(\theta_{IX}, \theta_{IY}, \dots, \theta_{ZZ})$---but our algorithm can easily simulate more complex models describing (e.g.) crosstalk or spatial inhomogeneity. (b-c) show that, for one representative choice of $\vec{\theta}$ (see main text), coherence can significantly enhance or suppress each individual qubit's marginal error probability ($\varepsilon$).  We quantify this effect by each qubit's coherent enhancement $\chi_{\varepsilon} \equiv \varepsilon_{\textrm{coh}} / \varepsilon_{\textrm{stoc}}$, the ratio of that qubit's $\varepsilon$ in the coherent noise model ($\varepsilon_{\textrm{coh}}$) to its value in a noise model in which the errors are stochastic but otherwise identical ($\varepsilon_{\textrm{stoc}}$). (b) shows $\chi_{\epsilon}$ for each qubit in a distance-11 code. (c) shows the mean $\chi_{\epsilon}$ for each kind of qubit for each distance $d$. (d) shows how another relevant quantity, the total expected number of syndrome bit flips ($\omega$), depends on $\vec{\theta}$ for $d=11$.  To leading order, $\omega_{\textrm{coh}} = \vec{\theta}^T S_\omega \vec{\theta} + O(|\vec{\theta}|^3)$, and our algorithm was used to compute $S_\omega$.  We also used it to capture $\omega$'s dependence on stochastic errors by computing a vector $\vec{v}_{\omega}$ such that $\omega_{\textrm{stoc}} = \sum_P v_{\omega, P} \theta_P^2$.  The bar chart in (d) compares $\vec{v}_{\omega}$ to the diagonal of $S_{\omega}$ (which encodes the dependency of $\omega_{\textrm{coh}}$ on $\theta_P^2$), showing that coherent addition and cancellation of errors strongly impacts $\omega$ for some values of $\vec{\theta}$. (e) demonstrates that the precise size of the coherent enhancement or suppression of $\omega$ depends strongly on $\vec{\theta}$, by plotting $\omega_{\textrm{coh}}$, $\omega_{\textrm{stoc}}$ and $\chi_\omega =\omega_{\textrm{coh}}/\omega_{\textrm{stoc}}$ for a one-dimensional family of $\vec{\theta}$ parameterized by a scalar $\xi$ (details in main text).}
\label{fig:main_results}
\end{figure*}

\customsection{Our contribution.} We present and demonstrate an efficient classical algorithm for simulating $n$-qubit Clifford circuits with near-arbitrary small Markovian errors.  Our only requirements on the noise are that it be \emph{sparse} and \emph{small}.  More technically, the error process for each layer of the circuit being simulated must be well-approximated by the exponential of a Lindbladian \cite{Blume-Kohout2022-ln, Malekakhlagh2025-ar} that is sparse in the elementary error generator representation of Ref. \cite{Blume-Kohout2022-ln}, and the sum of all layers' error Lindbladians must be at most $\mathcal{O}(1)$. Our algorithm is motivated by the insight that these constraints, while fundamental, permit many useful and interesting simulations.  For example, any error model in which each gate causes errors on at most $\mathcal{O}(1)$ qubits (e.g., its target qubits and a small neighborhood) is necessarily sparse.  Concomitantly, circuits implementing quantum algorithms or quantum error correction are most interesting when they experience relatively few errors.

We demonstrate the power of our algorithm by studying how coherent errors impact three interesting classes of large circuits.  First, as a warmup designed to confirm our algorithm's accuracy, we simulate creating and then uncreating a 100-qubit Greenberger-Horne-Zeilinger (GHZ) state \cite{Bouwmeester1999-wz} using noisy gates with coherent errors.  This is a standard task in quantum sensing and metrology, because the GHZ state is hypersensitive to external fields \cite{Bollinger1996-cb, Bao2024-bi, Proctor2018-lk}.  Although 100 qubits greatly exceeds the capability of general-purpose vector-state simulations, this example's symmetries permit an analytic solution, so we can confirm the accuracy of our algorithm (which does not leverage the symmetry) by comparing it to the analytic calculation.  Second, we simulate and calculate the infidelity of 225-qubit random Clifford circuits of the type used in randomized benchmarking \cite{Magesan2011-hc, Proctor2019-gf, Hines2024-qe}, with depth up to 8192. We tune how fast these circuits scramble errors by varying their composition, and observe how the impact of coherent errors on infidelity grows as scrambling gets slower.  Finally, we tackle one of the most timely open questions in quantum computing by simulating coherent gate errors in syndrome extraction circuits for high-distance ($d=3,5,7,9,11$) surface codes \cite{Google-Quantum-AI-and-Collaborators2025-ad} involving up to 241 qubits (see Fig.~\ref{fig:main_results}). By computing each physical qubit's probability of experiencing an error, and comparing it to the error rate that would be observed for depolarizing gate errors with the same fidelity, we show that coherent gate errors add up constructively on some qubits and increase their error rates by up to $\sim 3\times$, but interfere destructively on other qubits to cause as little as $\sim 0.33\times$ as much error.

\begin{figure}[t!]
\centering
\includegraphics[width=8.5cm]{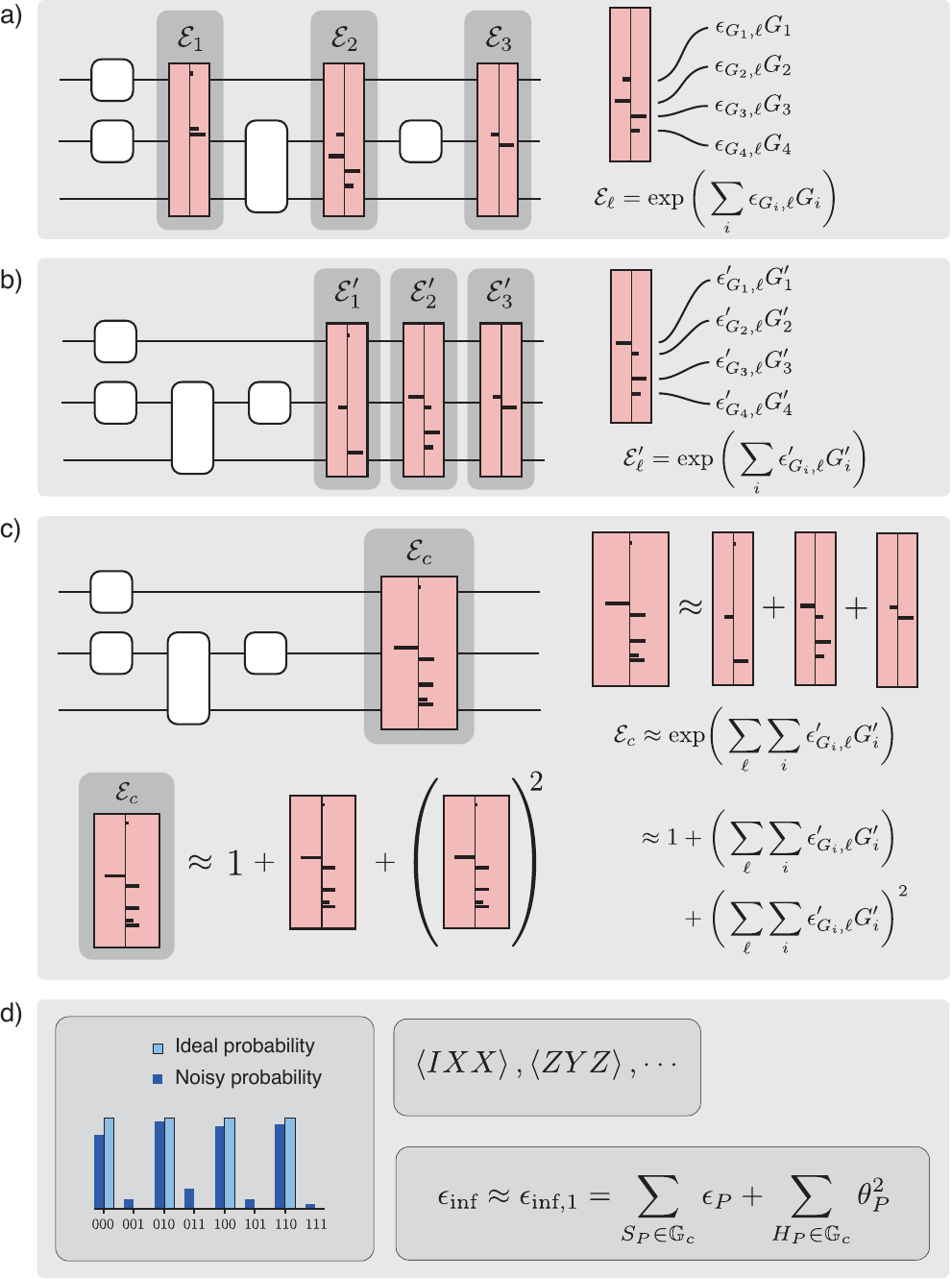}
\caption{\textbf{Efficient approximate simulation of noisy Clifford circuits.} Our algorithm for simulating a noisy Clifford circuit subject to sparse Lindbladian noise, including small coherent errors. This algorithm is polynomial in the number of qubits $n$, accurately models the addition and cancellation of coherent errors, and it enables efficiently approximating noisy outcome probabilities (strong simulation), Pauli expectation values, or the noisy circuit's process fidelity. }
\label{fig:schematic}
\end{figure}

\customsection{Simulation algorithm.} Our algorithm is illustrated in Fig.~\ref{fig:schematic}. Its efficiency stems from two ideas originating in Ref.~\cite{Blume-Kohout2022-ln}: \emph{sparse} representations of $n$-qubit error processes and \emph{perturbative} expansions of small error channels.  By combining them, we can approximate the effects of general errors using only symbolic and extremely fast $n$-qubit Pauli group and stabilizer state calculations. Applying standard efficient representations of Clifford unitaries \cite{Gidney2021-ef} to sparse representations of noise processes completely circumvents any need for multiplying exponentially large matrices. Our algorithm only works when the errors are small---a necessary limit, since efficient classical simulation of circuits containing many large non-Clifford unitaries (e.g.~large coherent errors) would imply BQP $=$ BPP.  It nonetheless enables exploration of many interesting scientific questions about the effects of coherent errors.

We consider noisy Clifford circuits of width $n$ and depth $d$---i.e., $n$-qubit circuits comprising $d$ consecutive layers of Clifford unitaries, interleaved with \emph{sparse Lindblad noise} processes $\mathcal{E}_i$, as illustrated in Fig.~\ref{fig:schematic}a.  Each layer's noise process can be written as $\mathcal{E}_i = \exp(\sum_{G \in \mathbb{S}_{i}} \epsilon_{i,G} G)$, where $\mathbb{S}_{i}$ is a small set containing just $\kappa = \textrm{poly}(n)$ elements of the $\mathcal{O}(16^n)$-element basis for $n$-qubit Lindbladians introduced in Ref.~\cite{Blume-Kohout2022-ln}.  In this basis, an arbitrary Lindbladian is written as a sum of Hamiltonian ($H$), Pauli stochastic ($S$), Pauli correlation ($C$), and active ($A$) ``elementary error generators''.  Each one is indexed by one ($H$ and $S$) or two ($C$ and $A$) $n$-qubit Pauli operators.  We leverage three key features of this basis:
\begin{enumerate}
\item Many noise processes are exactly or approximately sparse ($\kappa$ is small) in this representation \cite{Blume-Kohout2022-ln}.
\item Many properties of these basis elements can be computed symbolically using analytic formulae (see Methods) together with standard properties of the Pauli and Clifford group. They include commutators and products of any two basis elements, conjugation of a basis element by a Clifford superoperator, and the expectation values $\textrm{Tr}\left(\ket{x}\bra{x}G[\ket{\psi}\bra{\psi}]\right)$ and $\textrm{Tr}\left(PG[\ket{\psi}\bra{\psi}]\right)$, where $\psi$ is any stabilizer state, $P$ any Pauli operator, and $G$ any basis element. 
\item Conjugation by any Clifford superoperator preserves sparsity of $\mathcal{E}_i$, because $G \to \pm G'$ where $G'$ is also a basis element.
\end{enumerate}

The first step in our algorithm is to propagate all error channels to the end of the circuit, $\mathcal{E}_i \to \mathcal{E}_i'$, as shown in Fig.~\ref{fig:schematic}b, giving a \emph{circuit error process} $\mathcal{E}$ that is equal to the product of the circuit's constituent propagated error processes $\mathcal{E}  = \mathcal{E}_d'\cdots\mathcal{E}_1'$. The properties of the basis described above allow this to be done exactly, efficiently, and without changing the error channels' sparsity. The only thing that changes is the identity of the errors (e.g., a few-qubit error might become a many-qubit error) and possibly their signs. Next, we apply perturbative expansions (the Baker–Campbell–Hausdorff [BCH] formula \cite{BLANES2004135} and a Taylor expansion) to compute approximate representations of $\mathcal{E}$ (Fig.~\ref{fig:schematic}c) and then evaluate properties of the circuit outcome distribution (Fig.~\ref{fig:schematic}d). 

We combine $\mathcal{E}_i$'s sparse Lindbladians into a single Lindbladian for $\mathcal{E}$ (whose exponential is $\mathcal{E}$) using the $k^{\textrm{th}}$ order BCH approximation. This approximation contains $\mathcal{O}\left((d\kappa)^k\right)$ terms, each of which is a basis element multiplied by a coefficient (error rate).  Each of those terms can be computed efficiently. We are not done yet, though, because answering most interesting scientific questions requires an approximation to $\mathcal{E}$, not its Lindbladian. We construct this approximation using an $l^{\textrm{th}}$-order Taylor expansion of $\mathcal{E}$. This approximation to $\mathcal{E}$ is a linear combination of $\mathcal{O}\left((d\kappa)^{kl}\right)$ basis elements, each of which can be computed efficiently. The expansions we use are valid when the total error in the circuit $\epsilon = \sum_{i,G\in \mathbb{S}_{i}}|\epsilon_{i,G}|$ is small.

These steps yield efficient approximate representations of $\mathcal{E}$ and $\mathcal{E}$'s Lindbladian.  They can be interrogated directly to learn useful properties of the noisy circuit. Efficiently compututable properties include the probability of many-qubit errors, the degree to which coherent errors add or cancel within the circuit, and approximations to process fidelity and diamond norm error \cite{Madzik2022-jh}. For example, the process fidelity of $\mathcal{E}$ is well-approximated by $F = \epsilon_{S} + \epsilon_{H}$ where $\epsilon_{S}$ is the total rate of all Pauli stochastic error terms and $ \epsilon_{H}$ is the sum of the squares of the Hamiltonian (coherent) terms in $\mathcal{E}$'s 1\textsuperscript{st} order Lindbladian approximation \cite{Madzik2022-jh}. But the most common desiderata for noisy circuit simulation are the probabilities of measurement outcomes, and the expectations of observables. The probability of observing any bit string $x$, or the expectation value of any Pauli operator $P$, can be computed from our $\mathcal{O}\left((d\kappa)^{kl}\right)$-term expansion of $\mathcal{E}$ in our Lindbladian basis, by applying our efficient formulae for $\textrm{Tr}\left(\ket{x}\bra{x} G[\ket{\psi}\bra{\psi}]\right)$ or $\textrm{Tr}\left(P G[\ket{\psi}\bra{\psi}]\right)$ to each term, where $\psi$ is the state that the circuit would produce in the absence of errors. 

The running time of our simulation algorithm is $\mathrm{poly}(n)$ for any fixed $k$ and $l$, as long as the noise model's sparsity $\kappa$ and the circuit depth $d$ are $\textrm{poly}(n)$. Moreover, the exponents in the polynomial are small, implying practical feasibility, for relevant scenarios. For example, in many experimentally-relevant noise models like localized coherent errors, $\kappa$ grows linearly with $n$. Small values of $k$ and $l$ enable many scientifically interesting simulations (e.g., the examples below), because a 1\textsuperscript{st}-order BCH expansion ($k=1$) captures to leading order how coherent errors interfere constructively and/or destructively across an entire circuit, and 2\textsuperscript{nd} or 1\textsuperscript{st}  order Taylor expansions ($l=2,1$) capture the leading order contribution of those coherent errors to definite-outcome measurements (ones whose noise-free probabilities are 0 or 1) and indefinite-outcome measurements (all others), respectively. In the examples presented below, we were able to explore the impact of coherent errors in circuits with hundreds of qubits using simulations that are feasible (and in many cases take only a few seconds) on an ordinary laptop computer.

\begin{figure}[t!]
\centering
\includegraphics[width=8.6cm]{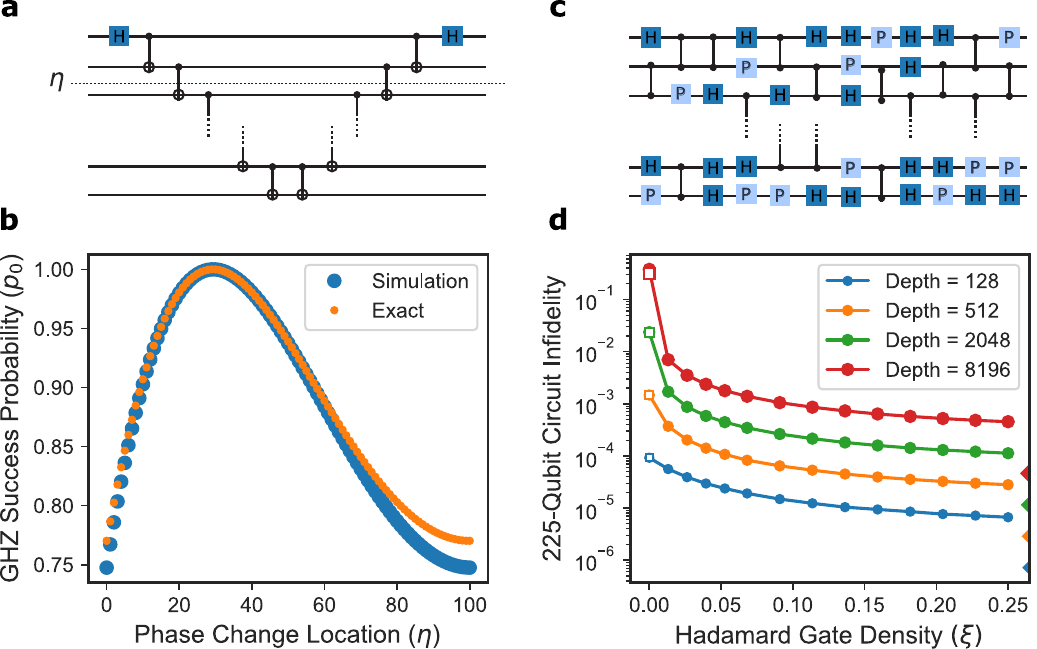}
\caption{\textbf{Coherent errors in many-qubit circuits}. We used our algorithm to study the effect of coherent errors in (a) a circuit that prepares a 100-qubit GHZ state and then unentangles the qubits, and (c) 225-qubit random circuits up to depth 8192. The GHZ state circuit was simulated under a simple coherent noise model with a tunable parameter $\eta$ (details in text) for which we could also analytically compute the correct results. We observe close agreement between our approximation and the exact, analytic answer, for the circuit's success probability---the probability all qubits return 0. We simulated the random circuits for a noise model with coherent phase errors, and we varied the circuit's scrambling power by varying the density of Hadamard gates within it. We find that the process infidelities of these circuits vary by orders of magnitude as Hadamard density varies, showing how coherent errors can combine even within random circuits.}
\label{fig:main_results_1}
\end{figure}

\customsection{Example applications.} 

\customsubsection{Simulating noisy GHZ state creation.---}To evaluate and demonstrate our algorithm's accuracy, we compute outcome probabilities for a 100-qubit circuit inspired by metrology, with coherent errors on each gate.  This circuit, shown in Fig.~\ref{fig:main_results_1}a, creates a 100-qubit GHZ state and then uncreates it.  If no errors occur, measuring each of the 100 qubits will yield 0, so the probability $p_0$ of this outcome is a simple measure for circuit success. We simulated a noise model in which each gate (including single-qubit idles) is followed by a coherent error that rotates the target qubit around the $z$-axis by either $-\theta$, for the first $\eta$ qubits, or $+\theta$, for the rest. This is a simple model for a linear chain of qubits in a spatially varying magnetic field, and it is convenient because we can compute $p_0$ exactly as $p_{0} = \cos^2 (\nicefrac{\theta_{\textrm{acc}}}{2})$, where $\theta_{\textrm{acc}}$ is the sum of all the phases accrued by each qubit in between the first and last gate on that qubit.  To simulate these circuits, we set $k=1$ and $l=2$ in our algorithm.

Fig.~\ref{fig:main_results_1}b shows our algorithm’s prediction for $p_{0}$ as a function of $\eta$ (blue circles), and compares it to the exact calculation (orange dots), for $\theta=10^{-4}$. They agree well, demonstrating that our algorithm can achieve good accuracy, at least for this analytically solvable problem, which is far beyond what vector-state simulations can feasibly simulate.

\vspace{0.1cm}
\customsubsection{Coherent error scrambling in random circuits.---}Next, we explore how coherent errors impact random circuits intended to \emph{scramble} errors.  Scrambling suppresses coherent interference.  It manifests most strongly in circuits comprising a sequence of unitary gates drawn randomly from a unitary 2-design \cite{Dankert2009-yt}. If an error process $\mathcal{E}$ is interleaved with these unitaries, then the effective impact of $\mathcal{E}$ is the same as a uniform depolarizing process, without any coherent addition or cancellation of errors even if $\mathcal{E}$ is actually coherent \cite{Dankert2009-yt, Gross2007-lp, Hashim2024-om}. But realistic implementations of scrambling circuits are not so simple. The random unitaries must be compiled into many consecutive ``native'' gates \cite{Magesan2011-hc, Proctor2019-gf} or replaced with random layers of those gates \cite{Proctor2019-gf, Hines2024-qe, Boixo2018-kp, Arute2019-mk}. So, in practice, error processes are interleaved with layers of native gates that do not immediately scramble them, leaving room for some coherent addition of errors.  We used our simulation algorithm to explore this effect.

Specifically, we simulated the impact of coherent errors on the fidelity of 225-qubit random circuits comprising up to 8196 layers of one- and two-qubit Clifford gates (Fig.~\ref{fig:main_results_1}c), and containing up to $1.8$ million gates.  The native gates were drawn at random from a 3-gate set: the controlled-$Z$ gate, the Hadamard gate, and the phase gate. We simulated the circuits under a noise model in which, after each layer of gates, each qubit experiences a coherent rotation by $\theta = 10^{-5}$ around the $z$-axis.  In the limit of perfect scrambling, this error process is randomized by the gates, and behaves (effectively) like weak depolarization.  But in the limit of no scrambling, where the error processes commute with the gates, the small $z$-axis rotations all add up coherently and have a much greater impact on the overall circuit fidelity.  We varied the circuits' scrambling power by varying the density of Hadamard gates ($\xi$), so that the circuits' scrambling power decreases as $\xi \to 0$.

Figure~\ref{fig:main_results_1}d shows our algorithm's approximation to the random circuits' process infidelities, for four representative circuit depths, versus $\xi$. When no Hadamard gates are included ($\xi=0$), there is no scrambling at all. The $z$-axis errors in our model commute with every gate, and add up constructively. Our algorithm correctly predicts this effect: for $\xi=0$ the circuits' process fidelities are exactly given by $\cos^{2n}(\nicefrac{d \theta}{2})$ (squares in Fig.~\ref{fig:main_results_1}d) which agrees closely with our algorithm's prediction. As we increase $\xi$ from 0 up to $0.25$, we observe a large decrease in circuit infidelity, by up to $1000 \times$ for the deepest circuits. Notably, even at the highest density of Hadamard gates, scrambling remains far from perfect---circuit infidelities remain approximately $10 \times$ larger than they would be in the theoretical limit of perfect scrambling (i.e., where each layer implements a random element of a unitary 2-design), which we calculated analytically and indicate by triangles on the right edge of Fig.~\ref{fig:main_results_1}d. These simulations therefore show and precisely quantify the significant degree to which coherent errors can add constructively even in random circuits.

\vspace{0.1cm}
\customsubsection{Coherent errors in surface code circuits.---}Finally, we used our algorithm to quantify the impact of coherent errors on syndrome extraction circuits for rotated surface codes with distance $d=3, 5, 7, 9$, and $11$ (Fig.~\ref{fig:main_results}a). We focused on how coherent errors combine within a single round of syndrome extraction, since we expect measurements of the syndrome qubits after each syndrome extraction round to prohibit significant coherent addition and cancellation of errors between rounds \cite{Beale2018-bz, Iverson2020-ym} (although our algorithm could be used to simulate multiple rounds of syndrome extraction). The data qubits are initialized in the $\ket{0}$ logical state (a $+1$ eigenstate of all code stabilizers and the logical $Z$ operator) and all syndrome qubits are initialized in $\ket{0}$. We simulate a noise model in which each CNOT and each Hadamard gate (regardless of which qubit[s] it targets) is followed by an arbitrary unitary error on the gate's target qubit[s]. This model is parameterized by 3 Hamiltonian rates for the Hadamard gates ($\theta_X$, $\theta_Y$, and $\theta_Z$), and 15 Hamiltonian rates for the CNOT gate ($\theta_{IX}$, $\theta_{IY}$, $\dots$, $\theta_{ZZ}$).  We represent these parameters as an 18-element vector $\vec{\theta} = (\theta_X,\theta_Y,\theta_Z,\theta_{IX},\theta_{IY},\cdots, \theta_{ZZ})$, and they correspond to twice the rotation angle around each axis \cite{Blume-Kohout2022-ln}. To quantify the impact of the errors' coherence, we also simulate an equivalent stochastic error model, where each Hamiltonian error with rate $\theta_P$ and Pauli axis $P$ is replaced with a stochastic $P$ error of rate $\theta_P^2$. It is parameterized by the vector $\vec{\epsilon} = (\theta_X^2,\theta_Y^2,\dots, \theta^2_{ZZ})$. 

We computed how coherent errors enhance or suppress the marginal probability of error on each qubit (Fig.~\ref{fig:main_results}b-c). We quantify this enhancement/suppression using the \emph{coherent amplification factor} $\chi$ of each qubit's marginal error rate $\varepsilon$. The coherent amplification of a quantity $x$ is $\chi_x \equiv x_{\textrm{coh}}/x_{\textrm{stoc}}$ where $x_{\textrm{coh}}$ is $x$'s value in our coherent error model and $x_{\textrm{stoc}}$ is its value in the equivalent stochastic model. The coherent amplification factor depends on the rotation angles $\vec{\theta}$ in the error model, so for Fig.~\ref{fig:main_results}b-c we chose specific parameter values:  $\theta_Z = \theta_{IX} = -\theta_{ZI} = 0.001$, with all other rates set to zero. For these model parameters, we find that coherent combination of errors causes an increase in some qubits' marginal error rates by $3.33\times$, while other qubits' marginal error rates are suppressed by $0.33 \times$. This reveals that coherent errors can have a much larger impact on syndrome extraction circuits than their gate fidelities suggest.

Our simulation algorithm enabled us to efficiently explore how properties of the syndrome extraction circuits depend on $\vec{\theta}$, because it computes an analytic approximation of the syndrome extraction circuit's Lindbladian as a function of the errors on the individual gates. In our coherent (resp., stochastic) model, the full circuit's approximate Lindbladian contains $\mathcal{O}(d^2)$ different Hamiltonian (resp., stochastic) basis elements, $\{H_Q\}$ (resp., $\{S_Q\}$) with $Q$ an indexing Pauli operator. The rate for each $H_Q$ (resp. $S_Q$) is given by $\theta_{Q} = \vec{s}_Q^T \vec{\theta}$ (resp., $\epsilon_{Q} = \vec{q}_Q^T \vec{\epsilon}$) for some integer-valued vector $\vec{s}_Q$ (resp., $\vec{q}_Q$). Our algorithm computes observable properties' dependence on these parameters, enabling us to compute \emph{sensitivities} of various quantities to coherent and stochastic errors.

Many interesting properties of noisy syndrome extraction circuits are (to leading order) quadratic in the $\theta_{Q}$ and linear in $\epsilon_Q$.  In the coherent model, $x$ is given by $x_{\textrm{coh}} = \sum_{Q,Q'}a_{Q,Q'} \theta_{Q}\theta_Q' + \mathcal{O}(\theta^3)$, while in the stochastic model it is given by $x_{\textrm{stoc}} = \sum_{Q}b_{Q} \epsilon_{Q} + \mathcal{O}(\theta^4)$, for some coefficients $a_{Q,Q'}$ and $b_Q$. Any such quantity can be expressed as $x_{\textrm{coh}} = \vec{\theta}^T S_x \vec{\theta}$ where $S_x = \sum_{Q,Q'} a_{Q,Q'} \vec{s}_Q\vec{s}_{Q'}^T$ is a matrix encoding $x$'s sensitivity to $\vec{\theta}$ and $x_{\textrm{stoc}} = \vec{v}^T_x \vec{\theta}$ where $\vec{v}_x = \sum_{Q} b_{Q} \vec{q}_Q$ is a vector encoding $x$'s sensitivity to $\vec{\epsilon}$. Examples of properties of this sort include the marginal probability of a bit flip on any syndrome qubit (as shown in Fig.~\ref{fig:main_results}b for a particular value of $\vec{\theta}$), the probability of each error in $\mathcal{E}$'s equivalent stochastic Pauli channel (the ``Pauli-twirl'' of $\mathcal{E}$ \cite{Hashim2024-om, Harper2023-pv}), and the marginal error probability on any data qubit. These sensitivity matrices allow us to, for example, calculate Fig.~\ref{fig:main_results}b for any $\vec{\theta}$ with almost no additional computational cost.

We constructed and examined the sensitivity matrix for the rate of syndrome bit flips ($\omega$) in the $d=11$ syndrome extraction circuit (Fig.~\ref{fig:main_results}d). It reveals that $\omega$ is strongly dependent on the structure of the coherent errors.  
In the bar chart of Fig.~\ref{fig:main_results}d, we compare $S_{\omega}$'s diagonal (which encodes $\omega_{\textrm{coh}}$'s sensitivity to $\theta_P^2$ terms) to $\omega$'s sensitivity to the equivalent stochastic errors, quantified by the vector $\vec{v}_{\omega}$ (as $\omega_{\textrm{stoc}} = \vec{v}_{\omega}^T\vec{\epsilon} = \sum_P v_{\omega,P}\theta_P^2$). These sensitivities differ substantially, which shows that there is significant coherent enhancement or suppression of $\omega$ (depending on $\vec{\theta}$'s value) due to coherent interactions of the same error parameters in different circuit locations. $S_{\omega}$ also contains substantial off-diagonal elements, which have no equivalent in the stochastic model. These off-diagonal elements show that $\omega$ is significantly enhanced or suppressed (again, depending on $\vec{\theta}$'s value) by coherent combination of two different errors ($\theta_{Q}$ and $\theta_{Q'}$) occurring at different locations in the circuit. 

To illustrate how the coherent enhancement of $\omega$ varies with $\vec{\theta}$, in Fig.~\ref{fig:main_results}e we plot $\omega_{\textrm{coh}}$, $\omega_{\textrm{stoc}}$, and their ratio $\chi_{\omega}=\omega_{\textrm{coh}}/\omega_{\textrm{stoc}}$ for a representative one-dimensional slice of our 18-dimensional parameter space, given by $\theta_Z = 0.001$, $\theta_{IX} = 0.001(\zeta -1)$ and $\theta_{ZI} = 0.001(1-\zeta)$. We find that $\chi$ varies from a little over 3 to just over 0.5---corresponding to substantial coherent enhancement and suppression of syndrome bit flip rates, respectively---as $\zeta$ varies. This shows how different kinds of coherent errors are more deleterious to syndrome extraction circuits than others. We conjecture this information could be used to inform gate calibration routines that aim to maximize quantum error correction performance.

\customsection{Discussion.} Quantum error correction is expected to be necessary for useful quantum computations \cite{Google-Quantum-AI-and-Collaborators2025-ad, Proctor2025-cd}, but understanding its behavior under coherent noise models has remained a largely open problem until now. Our simulation algorithm enables study and prediction of how small sparse Markovian errors impact error correction primitives including (but not limited to) syndrome extraction. We confirmed that coherent errors can coherently combine within surface code syndrome extraction circuits, increasing effective error rates by up to approximately $3\times$ relative to stochastic models with the same gate fidelities. 

Our simulation algorithm has much wider application than this single example, and it could be used to study many important and timely questions about quantum error correction. Intriguing possibilities including studying the effects of other kinds of errors (e.g., coherent $ZZ$ couplings or other crosstalk errors \cite{Sarovar2020-pz}) in quantum error correction, studying the effects of coherent errors on codes other than the surface code, and explorations of decoders that are optimized to coherent errors. One particularly exciting application is the estimation of logical qubit error rates and thresholds against realistic coherent (or mixed coherent and stochastic) errors. This would greatly extend existing results on surface code thresholds with coherent errors that are derived for specific kinds of coherent error (e.g., $z$-axis errors) and largely assume perfect syndrome extraction \cite{Bravyi2018-oi, Venn2020-of, Pato2025-qm, Venn2023-ap, Behrends2024-fm, Behrends2024-ok, Marton2023-il, Pataki2024-gx, Suzuki2017-jh}. We propose that this could be achieved by using the techniques we demonstrated here to compute an effective Pauli channel for one or more consecutive syndrome extraction cycles containing coherent errors, and then applying existing techniques, for extremely fast simulation of stochastic Pauli noise in many rounds of syndrome extraction \cite{Gidney2021-ef}, to compute logical qubit error rates from that Pauli channel.

Another promising application, beyond what was explored here, is learning general Lindblad noise models \cite{Malekakhlagh2025-ar, Blume-Kohout2022-ln} for many-qubit systems from data.  Tomographic protocols for learning error models rely on fitting those models to data \cite{Blume-Kohout2017-no, Madzik2022-jh, Malekakhlagh2025-ar}, which in turn requires efficient simulation of those models. Our simulations enable efficient simulation of sparse noise models, and therefore open the door to efficiently fitting sparse Lindblad models to experimental data.  They have the potential to enable an \emph{efficient} version of gate set tomography \cite{Blume-Kohout2017-no, Madzik2022-jh}, extending efficient Pauli noise learning methods \cite{Harper2023-pv, Harper2020-te, Harper2021-tm,Flammia2021-dn, Hines2025-jf} to sparse but otherwise general noise. 

Although our algorithm is specifically designed to simulate noisy Clifford circuits, it may be extendable to other restricted classes of circuit. For example, it might be possible to combine our methods with existing algorithms for practical (but exponentially scaling) simulation of Clifford circuits that contain a small number of $T$ gates, or other special circuit families \cite{Bravyi2016-hq, Huang2019-ma, Kocia2020-fm, Smith2023-mu, Park2024-nr, Bu2019-xg}. Finally, our methods can be used to simulate certain kinds of non-Markovian errors by promoting the rates of individual errors from real numbers to time-dependent stochastic processes. This would enable exploration of how non-Markovian errors impact many-qubit circuits, and could enable the efficient simulation of quantum error correction circuits with truly realistic noise models.

\begin{small}
\section*{Methods}

\customsubsection{Clifford circuits.---}Our algorithm simulates $n$-qubit circuits constructed from a set of layers $\mathbb{L} = \{\layer\}$, each of which performs an $n$-qubit Clifford unitary $U_\layer \in \textrm{SU}(2^n)$ in the absence of any errors. We usually assume that each layer $\layer$ is constructed from parallel applications of one- and two-qubit gates, although this is not necessary. Our methods can also be applied to circuits containing layers that implement Pauli-basis mid-circuit measurements (MCMs) without feed-forward, by employing the principle of deferred measurement, as long as errors in the MCM are represented by error channels before and after the error-free MCM.  However, for simplicity, we do not include MCMs in our notation.

\vspace{0.2cm}
\customsubsection{Elementary error generators.---} The elementary error generators (EEGs) are $n$-qubit superoperators that come in four types: Hamiltonian, stochastic, correlated, and active. They are defined, respectively, by their action on an $n$-qubit density operator $\rho$:
\begin{align}
    \eg{H}_{P}\left[\rho\right]&=-i\left[P,\rho\right] ,\\
    \eg{S}_{P}\left[\rho\right]&=P\rho P- \rho, \\
    \eg{C}_{P,Q}\left[\rho\right]&=P\rho Q+Q\rho P-\frac{1}{2}\left\{\left\{P,Q\right\},\rho\right\},\\
    \eg{A}_{P,Q}\left[\rho\right]&=i\left(P\rho Q-Q\rho P+\frac{1}{2}\left\{\left[P,Q\right],\rho\right\}\right).
\end{align}
Here $P,Q \in \mathbb{P}^*$, where $\mathbb{P}^*$ is the set of $n$-qubit Pauli operators without signs and excluding the identity (e.g., $\mathbb{P}^* =\{X,Y,Z\}$ for $n=1$), and $Q>P$ according to some indexing of $\mathbb{P}^*$. There are $4^n(4^n-1)$ EEGs and we denote the set of all EEGs by $\mathbb{G}$, which is a basis for trace-preserving superoperators. Throughout the Methods, it will also be notionally convenient to let 
\begin{align}
\eg{G}(H,P,P) &\equiv H_P,   \\
\eg{G}(S,P,P) &\equiv S_P,  \\ 
\eg{G}(C,P,Q) &\equiv C_{P,Q}, \\
\eg{G}(A,P,Q) &\equiv A_{P,Q}.
\end{align}

\vspace{0.2cm}
\customsubsection{Representing error processes with sparse Lindbladians.---}A noise process on $n$ qubits is typically modelled by an $n$-qubit completely positive and trace preserving (CPTP) map $\mathcal{E}$, which is often represented as a dense $4^n \times 4^n$ matrix (e.g., in the Pauli basis) \cite{Hashim2024-om}. The EEGs enable an alternative parameterization of $\mathcal{E}$, as any small error map $\supop{E}$ (i.e., $\supop{E} \approx \id$) can be written (or very well approximated) as \cite{Blume-Kohout2022-ln}
\begin{equation}
\supop{E}= \exp(\mathcal{G}),
\end{equation}
where 
\begin{equation}
\mathcal{G} = \sum_{G\in\mathbb{G}} \epsilon_G\eg{G}.
\end{equation}
We will refer to $\epsilon_G$ as the rate of EEG $G$, although note that $\epsilon_G$ can be negative for some $G$ (e.g., Hamiltonian EEGs).

EEGs are a particularly useful basis in which to represent error maps because most physically realistic error processes on $n$ qubits are approximately \emph{sparse} (when $n \gg 1$) in this representation \cite{Blume-Kohout2022-ln}, i.e.,
\begin{equation}
\mathcal{E} \approx \exp\left(\sum_{G \in \mathbb{S}} \epsilon_G \eg{G}\right),
\end{equation}
where $\mathbb{S}$ is a small subset of $\mathbb{G}$.

\vspace{0.2cm}
\customsubsection{Sparse Lindblad noise models.---}Our simulation algorithm is designed for simulating sparse Lindblad noise models. A sparse Lindblad noise model describes the action of layer $\layer$ by a CPTP map 
\begin{equation}
\Lambda_\layer = \mathcal{E}_\layer \mathcal{U}_\layer
\end{equation}
with the layer's \emph{error map} $\mathcal{E}_\layer$ given by
\begin{equation}
\mathcal{E}_\layer = \exp\left(\sum_{\eg{G} \in \mathbb{G}_\layer}\epsilon_{\eg{G},\layer} \eg{G}\right), \label{eq:error_map}
\end{equation}
where $\mathbb{G}_\layer \subseteq \mathbb{G}$ is a $\kappa_{\layer}$-sized subset of the EEGs. Informally, the noise model is sparse if $\kappa_{\layer}$ is small for all layers in our circuits. The set $\mathbb{G}_\layer$ contains all those EEGs for which the model predicts non-zero rates, in layer $\layer$, and $\epsilon_{\layer,\eg{G}}$ is the rate that the model predicts for EEG $\eg{G}$ in $\mathbb{G}_\layer$. Our EEGs can also predict errors in the initial state preparation and final measurement, as error maps, with the form of Eq.~\eqref{eq:error_map}, after and prior to these operations, respectively. However, we do not explicitly denote these error maps anywhere herein, for notational simplicity.

\vspace{0.2cm}
\customsubsection{Conjugating basis elements by Clifford unitaries.---}Sparsity is preserved by any Clifford circuit, in the sense that the overall error map that the model predicts for any Clifford circuit is also sparse in the EEG basis. Conjugating any EEG $\eg{G}$ by the superoperator $\mathcal{U}$ for a Clifford unitary $U$ (i.e., $\mathcal{U}[\rho] = U\rho U^{\dagger}$) simply transforms $\eg{G}$ into a superoperator $\eg{G}_U$ where either $G_U$ or $-G_U$ is in $\mathbb{G}$. Specifically,
\begin{equation}
\label{Eqn:Propagation}
G_U \equiv \mathcal{U} G(T, P, Q) \mathcal{U}^{\dagger} = \gamma(G(T,P,Q),U) G(T, P_U , Q_U),
\end{equation}
where $P_U \in \mathbb{P}^*$ is defined by 
\begin{equation}
U^{\dagger} P U = s_{U,P} P_U
\end{equation}
with $s_{U,P} \in \{\pm 1\}$, and $ \gamma(G(T,P,Q),U)  \in \{\pm 1\}$. These $\gamma$ are given by 
\begin{align}
\gamma(S_P,U)&=1, \\
\gamma(H_P,U)&=s_{U,P}, \\
\gamma(A_{P,Q},U)&=s_{U,P}s_{U,Q}, \\
\gamma(C_{P,Q},U)&=s_{U,P}s_{U,Q}.
\end{align}
Derivations of these equations are provided in Supplemental Note~\ref{sm:eegs}. Therefore, $\eg{G}_U$ depends only on how $\eg{G}$'s indexing Pauli operator(s) transform under conjugation by $U$, which can be efficiently computed using the symplectic representation of $U$ \cite{Gidney2021-ef, Aaronson2004-ab}.  For example, $\supop{S}^{\dagger} \eg{H}_X \supop{S} = -\eg{H}_Y$ and $\supop{S}^{\dagger} \eg{S}_X \supop{S} = \eg{S}_Y$ where $\mathcal{S}[\rho]=S\rho S^{\dagger}$ is the superoperator for the phase gate $S$, where $S=\exp(-iZ/4)$ and $S^{\dagger}XS=-Y$. 

\vspace{0.2cm}
\customsubsection{Algorithm step 1: Error channel propagation.---}We now describe the first step in our simulation algorithm (Fig.~\ref{fig:schematic}). Our noise model $\mathcal{N}$ predicts that circuit $\circuit$'s superoperator is
\begin{equation}
\Lambda_\circuit = \mathcal{E}_{\layer_d}\mathcal{U}_{\layer_d} \cdots \mathcal{E}_{\layer_2}\mathcal{U}_{\layer_2}\mathcal{E}_{\layer_1}\mathcal{U}_{\layer_1},
\label{eq:lambda_circuit}
\end{equation}
where each error map is given by Eq.~\eqref{eq:error_map}. All the components in Eq.~\eqref{eq:lambda_circuit} are $n$-qubit superoperators, but they are described in an efficient representation. The first step in our simulation methods is to propagate all of the error maps to the end of the circuit, using only efficient operations on our efficient representation. Letting $A_{d:i} = A_{\layer_d}\cdots A_{\layer_{i+1}}A_{\layer_{i}}$ for any operator sequence $A_{\layer_1}$, $A_{\layer_2}$, $\dots$, $A_{\layer_d}$, we can write $\Lambda_\circuit$ as 
\begin{equation}
\Lambda_\circuit = \mathcal{E}_\circuit \mathcal{U}_\circuit 
\end{equation}
where $\mathcal{U}_\circuit = \mathcal{U}_{d:1}$, and $\mathcal{E}_{\circuit}$ is, by definition, $c$'s error map. This error $\mathcal{E}_\circuit = \mathcal{E}_{d:1}'$ can be written as
\begin{equation}
   \mathcal{E}'_{\layer_i}=  \mathcal{U}_{d:i+1}  \mathcal{E}_{\layer_i}  \mathcal{U}_{d:i+1}^{\dagger}.
\end{equation}
Now, each $\mathcal{E}'_{\layer_i}$ is given by
\begin{align}
\mathcal{E}'_{\layer_i} &= \exp\left(\sum_{G(T,P,Q) \in \mathbb{G}_{\layer_i}} \gamma(G(T,P,Q),U_{d:i+1}) \epsilon_{G,\layer_i} G(T, P_{U_{d:i+1}}, Q_{U_{d:i+1}}) \right) \notag \\
&= \exp\left(\sum_{G \in \mathbb{G}'_{\layer_i}} \epsilon_{G,\layer_i}' G \right).\label{eq:Edash}
\end{align}
Here, $\mathbb{G}'_{\layer_i}$ and each error rate $\epsilon_{G,\layer_i}'$ is an efficient representation of $\mathcal{E}'_{\layer_i}$. Furthermore, it can be computed for all $d$ layers using $\mathcal{O}(\kappa(n) d^2)$ computations that each consist of calculating how some Pauli operator is transformed by a Clifford unitary $U$. Each such computation is achievable by simple manipulations of $2n \times 2n$ matrices over $\{0,1\}$ and length $2n$ vectors (encoding $U$ and $P$) \cite{Gidney2021-ef, Aaronson2004-ab}.

\vspace{0.2cm}
\customsubsection{Algorithm step 2: Perturbative expansions.---}The next step of our algorithm is two perturbative approximations: applying the BCH formula and a Taylor expansion. We efficiently compute an approximate formula for $\mathcal{E}_{\circuit}$'s Lindblad representation using the BCH formula, from our efficiently-computable formula for $\mathcal{E}_{\circuit}$ [Eq.~\eqref{eq:Edash}]. $\mathcal{E}_{\circuit}$ has an exact decomposition into EEGs 
\begin{equation}
\mathcal{E}_{\circuit}  =\exp\left( \sum_{G \in \mathbb{G}} \epsilon_{\circuit,G}G \right),
\end{equation}
and we can compute an efficient \emph{approximation} to this decomposition by combining all the error channels $\mathcal{E}_{\layer_i}'$ that appear in Eq.~\eqref{eq:Edash} into a single error channel $\tilde{\mathcal{E}}_{\circuit,k}$. We do so using the $k$th order BCH formula: $\mathcal{E}_{\circuit}   \approx \tilde{\mathcal{E}}_{\circuit,k} = \exp(\sum_k \Omega_k)$ where $\Omega_1$ is simply the sum of each layers transformed error generators, i.e.,
\begin{align}
\label{Eqn:EOCPreBCH}
\Omega_1 =  \sum_{i=1}^{d}\sum_{G \in \mathbb{G}_{\layer_i}'}\epsilon_{G}'G,
\end{align}
and $\Omega_k$ for $k>1$ consists of $\mathcal{O}([d \kappa(n)]^k)$ terms each of size $\mathcal{O}(\epsilon^k)$, computed from the $(k-1)$-nested commutators of all the EEGs in all the layers, where $\epsilon$ bounds the size of each error rate (i.e., $\epsilon \geq \max_{\layer,G}|\epsilon_{G,\layer}|$).

Computing each term in $\Omega_k$ for $k \geq 2$ requires formulae for the commutators of any two EEGs, which we provide in Sec.~\ref{sec:commutators} of the Supplemental Material. Therefore, the error channel $\tilde{\mathcal{E}}_{\circuit,k}$ is represented by at most  $\mathcal{O}([d \kappa(n)]^k)$ different EEGs $\mathbb{G}_{\circuit,k}$ each associated with a rate $\epsilon_{\circuit,G,k}$, and both $\mathbb{G}_{\circuit,k}$ and $\{\epsilon_{\circuit,G,k}\}$ are efficient to compute for any constant $k$, resulting in an approximation of $\mathcal{E}_{\circuit}$ by an efficient representation of $\tilde{\mathcal{E}}_{\circuit,k}$ as
\begin{equation}
 \mathcal{E}_{\circuit} \approx \tilde{\mathcal{E}}_{\circuit,k}  =\exp\left( \sum_{G \in \mathbb{G}_{\circuit,k}} \epsilon_{\circuit,G,k}G \right).
\end{equation}

Whenever we require a representation of $\mathcal{E}$ rather than $\mathcal{E}$'s Lindbladian (for most purposes this is necessary), we apply an $l$th order Taylor expansion to our approximation for $\mathcal{E}$:
\begin{equation}
 \mathcal{E}_{\circuit} \approx 1 + \sum_{G \in \mathbb{G}_{\circuit,k}} \epsilon_{\circuit,G,k}G  + \left(\sum_{G \in \mathbb{G}_{\circuit,k}} \epsilon_{\circuit,G,k}G\right)^2 + \dots
\end{equation}
We can express our approximation in the EEGs basis using analytic formulae for the product of any two EEGs. Note, however, that for some quantities we do not have to explicitly compute this Taylor expansion---e.g., we do not need to do so to compute an approximate to the circuit's process fidelity.

\vspace{0.2cm}
\customsubsection{Computing a circuit's process fidelity.---}From our approximation for $\mathcal{E}_c$, above, we can compute many different properties of our noisy circuits. One interesting property is $\mathcal{E}_c$'s \emph{process infidelity} (or fidelity). To leading order in stochastic and Hamiltonian error rates, this is given by
\begin{equation}
\label{eqn:fidelity_est}
 \epsilon_{\textrm{inf}} \approx   \epsilon_{\textrm{inf},1} = \sum_{S_P \in \mathbb{G}_c} \epsilon_{P} + \sum_{H_P \in \mathbb{G}_c} \theta_P^2
\end{equation}
where $\{S_P \in \mathbb{G}_c\}$ and $\{H_P \in \mathbb{G}_c\}$ are the sets of all stochastic EEGs in $\mathbb{G}_c$, and $\epsilon_P \equiv \epsilon_{c,S_P}$ and $\theta_P \equiv \epsilon_{c,H_P}$ are the rates of $S_P$ and $H_P$ EEGs in the circuit's error generator, respectively. For $k=1$, this can therefore be computed with $d\kappa(n)$ operations, although it contains terms that are quadratic in the Hamiltonian rates.

\vspace{0.2cm}
\customsubsection{Computing circuit outcome probabilities.---}Strong simulation of a circuit $c$ under noise model $\mathcal{N}$ means computing $\mathcal{N}$'s prediction for the probability that $c$ outputs the bit string $x$ for any given $x$. This is given by
\begin{equation}
\tilde{p}_x = \textrm{Tr}(\ket{x}\bra{x}\Lambda_\circuit[\ket{0\cdots}\bra{0\cdots}]) = \textrm{Tr}(\ket{x}\bra{x}\mathcal{E}_{\circuit}[\ket{\psi}\bra{\psi}]) ,
\end{equation}
where $\Lambda_c$ is defined in Eq.~\eqref{eq:lambda_circuit}, and $\ket{\psi}=U\ket{0\cdots}$. We can approximately compute $\tilde{p}_x$ from our Taylor expansion approximation to $\mathcal{E}_c$, giving
\begin{equation}
\tilde{p}_x =  p_x + \sum^{k}_{j=1}\textrm{Tr}(\ket{x}\bra{x} \beta_j[\ket{\psi}\bra{\psi}])  + \mathcal{O}(\epsilon^{k+1}),
\end{equation}
where $\beta_j$ consists of $d \choose j$ products of $j$ elements of $\mathbb{G}_{\circuit} = \cup_i\mathbb{G}_{\layer_i}'$ multiplied by their corresponding error rates $\epsilon'_{G,\layer_i}$, and 
\begin{equation}
p_x =|\braket{\psi}{x}|^2.
\end{equation}
In particular, the first-order approximation for $\tilde{p}_x$ (i.e., $k=l=1$) is given by 
\begin{equation}
    \tilde{p}_x =  p_x + \frac{1}{2^{\zeta(\psi)}} \sum_{G \in \mathbb{G}_\circuit }  \alpha(\psi, G, x) \epsilon_{\circuit,G,1}  + \mathcal{O}(\epsilon^2)
\end{equation}
where $\zeta(\psi) \in \{1,2,\dots,n\}$ is the number of bits on which $\psi$'s distribution is uniform, 
\begin{equation}
\alpha(\psi, G, x) = 2^{\zeta(\psi)}\textrm{Tr}(\ket{x}\bra{x}G[\ket{\psi}\bra{\psi}]) \label{eq:alpha}
\end{equation}
is a factor that specifies which errors affect $\tilde{p}_x$ and by how much, and 
\begin{equation}
\epsilon_{c,G,1}=\sum_{i \in \mathbb{B}(G)}\epsilon_{G,\ell_i}'
\end{equation}
where $\mathbb{B}(G)$ is the indices of the layers in which $G \in \mathbb{G}_{\ell_i}'$. That is, $\epsilon_{c,G,1}$ is computed by summing up the transformed error rates for all the EEGs that occurred in the circuit and that become $G$ when pushed to the end of the circuit. Each $\alpha(\psi, G, x)$ is in $\{0, \pm 2, \pm 4\}$ and can be computed efficiently, using the formula that we provide in the Supplemental Note~\ref{sm:eegs}. Furthermore, there are only $|\mathbb{G}_\circuit| = \mathcal{O}(d\kappa(n))$ many $ \alpha(\psi, G, x)$ factors to calculate.

\vspace{0.2cm}
\customsubsection{Computing Pauli expectation values.---}Calculation of the expectation value for any Pauli operator $P$, denoted $\langle P \rangle $, follows the same logic as given above for computing the probability of a bit string. In particular,
\begin{equation}
    \langle P \rangle =  \langle P \rangle_{0}  +  \sum_{G \in \mathbb{G}_\circuit } \beta(\psi, G, P) \epsilon_{\circuit,G,1}  + \mathcal{O}(\epsilon^2)
\end{equation}
where $ \langle P \rangle_{0} = \textrm{Tr}(P\ket{\psi}\bra{\psi})$ and $\beta(\psi, G, P) = \textrm{Tr}(PG[\ket{\psi}\bra{\psi}])$. Efficient-to-evaluate formulae for all $\beta(\psi, G, P)$ are provided in Supplemental Note~\ref{sm:eegs}. Higher-order terms can also be calculated using the formulae for products of any two EEGs and the formulae for $\beta(\psi, G, P)$ (in particular, we compute $2\textsuperscript{nd}$-order terms for the syndrome extraction simulations presented in the main text).

\vspace{0.2cm}
\customsubsection{Random circuits simulation.---}Here we provide additional details for the random circuit simulations example presented in the main text. In this example, we explored how coherent errors add up in the random circuits consisting of i.d.d.~layers of one- and two-qubit Clifford gates. There are many possible classes of such random circuits, corresponding to different gate sets and different distributions from which each layer of gates is sampled. We fix the gate set to $\gate{CZ}$, the Hadamard gate $\gate{H}$, and the phase gate $\gate{P}$, with $\gate{CZ}$ gates only between connected qubits in some connectivity graph. Our layer sampling distribution is parameterized by the expected number of each gate: we use the `edgegrab’ algorithm described in \cite{Proctor2021-wt} to sample the two-qubit gates in a layer, and then one-qubit gates as i.i.d. samples from a distribution over $\gate{H}$ and $\gate{P}$. We simulated this class of random circuits on $n = 225$ qubits, arranged in a $15 \times 15$ square lattice, with depths 1, 2, 4, $\dots$, 8196. We varied the $\gate{H}$ gate density while fixing the $\gate{CZ}$ gate density (and therefore varying the $\gate{P}$ gate density). We simulated an error model in which, after every layer of gates, each qubit is subject to a local coherent $Z$-basis error with a rotation angle of $\theta =  10^{-5}$ (we observe similar results with larger $\theta$ and commensurately smaller circuit sizes $dn$).

\section*{Code and Data Availability}
Our simulation algorithm is implemented in open-source code, within the Python software \texttt{pyGSTi} \cite{Nielsen2020-rd}. This implementation uses \texttt{stim} \cite{Gidney2021-ef} for many of the Pauli and stabilizer state operations used by our algorithm. Python notebooks used to generate all data and results for the three examples presented in the main text will be released shortly. 

\section*{Acknowledgements}
This material was funded in part by the U.S. Department of Energy, Office of Science, Office of Advanced Scientific Computing Research. T.P. acknowledges support from an Office of Advanced Scientific Computing Research Early Career Award. Sandia National Laboratories is a multi-program laboratory managed and operated by National Technology and Engineering Solutions of Sandia, LLC., a wholly owned subsidiary of Honeywell International, Inc., for the U.S. Department of Energy's National Nuclear Security Administration under contract DE-NA-0003525. All statements of fact, opinion or conclusions contained herein are those of the authors and should not be construed as representing the official views or policies of the U.S. Department of Energy or the U.S. Government.

\bibliography{bibliography}
\end{small}
\onecolumngrid

\newpage
\section*{Supplemental Material}
\section{Additional Applications}
In this section we will provide an additional example of the application of our algorithm to the simulation of binary randomized benchmarking (BiRB) circuits with arbitrary Markovian noise.

\subsection{Simulating Binary-RB Under General Markovian Noise}
\begin{figure}[htbp]
    \centering
    \includegraphics[width=0.5\linewidth]{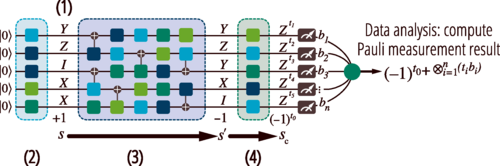}
    \caption{Each BiRB circuit consists of (1) sampling a random nonidentity Pauli operator, $s$, and performing a layer of single-qubit gates generating a +1 eigenstate of $s$, (2) $d$ layers of Clifford gates randomly sampled from some distribution $\Omega$, and (4) a final layer of single-qubit gates the Pauli operator resulting from propagation through the circuit ($s'$) into a tensor product of Z and I Pauli operators.}
    \label{fig:birb_ckt_structure}
\end{figure}

Binary randomized benchmarking (BiRB) is a highly scalable variant of randomized benchmarking capable of holistically benchmarking systems of dozens or even hundreds of qubits \cite{Hines2024-qe}. BiRB achieves this scalability by utilizing specially structured random quantum circuits without the use of motion reversal. Fig. \ref{fig:birb_ckt_structure} highlights the general structure of a BiRB circuit. The construction of each BiRB circuit consists of the following:

\begin{enumerate}
    \item Randomly sampling a nonidentity Pauli operator, $s$, and performing a layer of single-qubit gates which maps the input state to a $+1$ eigenstate of $s$.
    \item $d$ layers of Clifford gates sampled according to some distribution $\Omega$. Propagation through the randomly sampled Clifford layers has the effect of mapping the originally sampled Pauli operator $s$ to some new Pauli operator $s'$.
    \item A final layer of single-qubit gates mapping $s'$ to tensor-product of $Z$ and $I$ Pauli operators, $s_c$.
\end{enumerate}

\noindent Given a BiRB circuit, we'll refer to the expectation value of the final Pauli observable $\langle s_c \rangle$ as the \emph{energy} for that circuit. Using our algorithm one can compute approximations to the energies of an ensemble of BiRB circuits. This can enable, for example, studying the finer-grained structure of the distribution of the RB-data under different noise environments. In the context of traditional Clifford RB (CRB), for example, it is known that coherent errors strongly modify the distribution of survival probabilities for CRB circuits \cite{ball2016coherent}. 

As a demonstration of this application we simulated a complete BiRB experiment on $10$ qubits with a maximum circuit depth of $16$ under a noise model consisting of both random coherent and incoherent (including non-unital) errors. The gate set the circuits were constructed from consists of the single-qubit gates $\left\{R_X(\pm\frac{\pi}{2}),R_X(\pi), R_Y(\pm\frac{\pi}{2}), R_Y(\pi), R_Z(\pm\frac{\pi}{2}), R_Z(\pi)\right \}$, the two-qubit CPHASE gate, as well as single-qubit idle operations. For the noise model we randomly sampled a different set of CP-constrained local weight-1 Hamiltonian, stochastic and active error generator rates for each of the single-qubit gates and the idle operation. For the two-qubit CPHASE gate we instead sampled a random set of CP-constrained local weight-1 and 2 Hamiltonian, and weight-1 stochastic and active error generator rates.

\begin{figure}[h]
    \centering
    \includegraphics[width=.9\linewidth]{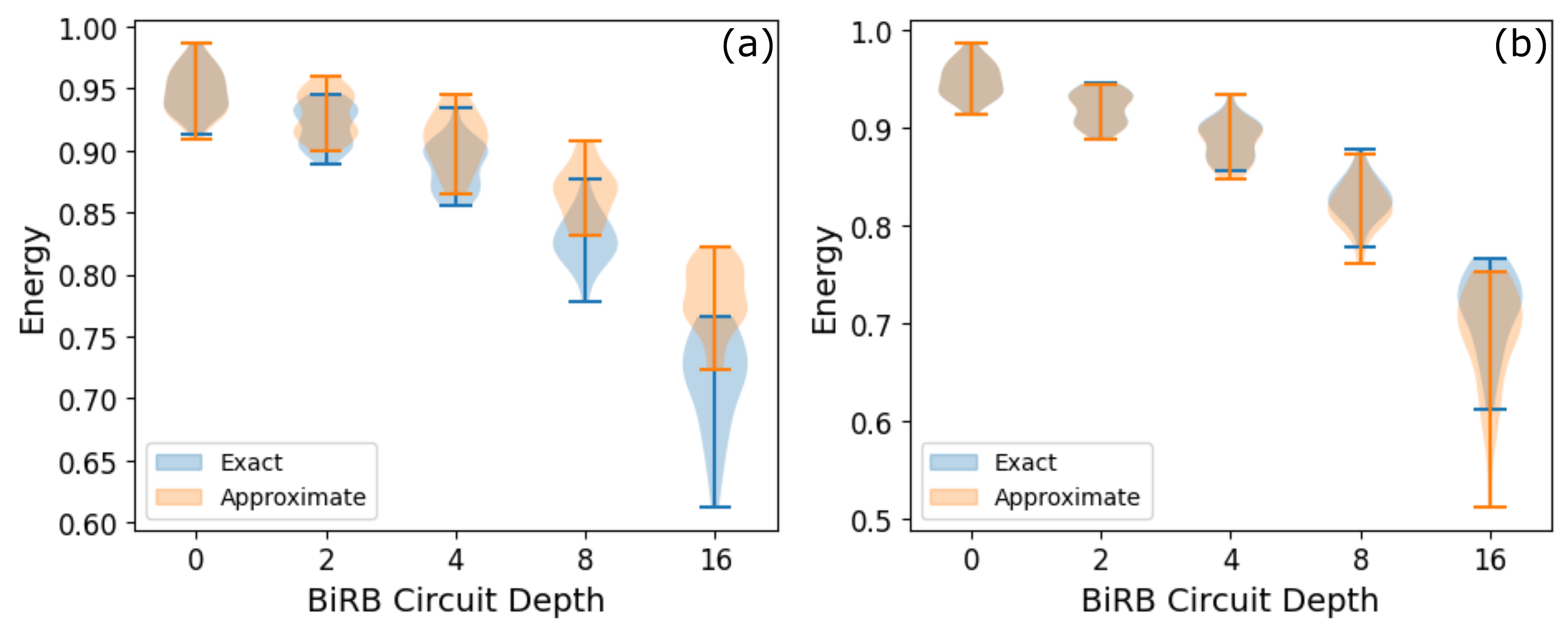}
    \caption{Comparison of the exactly computed BiRB circuit energies with those computed using error generator propagation using the first-order BCH approximation and (a) first-order, or (b) second-order, Taylor-series approximation. At higher circuit depths the first-order Taylor approximation fails to capture higher-order contributions the from constructive interference of certain error generators resulting in a systematic overestimation. The second-order Taylor approximation capture many of these contributions resulting in significantly improved simulation accuracy at higher depths.}
    \label{fig:birb_energy_exact_approx_combined}
\end{figure}

In Fig. \ref{fig:birb_ckt_structure} we compare the results of performing approximate error-generator-propagation-based simulation of the BiRB circuit energies, using the first-order BCH approximation and both first-order (Fig. \ref{fig:birb_energy_exact_approx_combined}(a)) and second-order (Fig. \ref{fig:birb_energy_exact_approx_combined}(b)) Taylor series approximation, to the exactly-computed energies. At short depths both the first- and second-order Taylor series approximations for the computation of the Pauli expectation values closely agree with both the average values and the distributions of the energies for the BiRB circuits. At greater circuit depths, however, the first-order Taylor approximation overestimates the energies, likely due to the failure to capture higher-order contributions from constructively interfering coherent errors. The second-order Taylor approximation on the other hand, which accounts for some of these higher-order contributions, agrees much more closely with the exact energy even at higher depths.

\begin{figure}[h]
    \centering
    \includegraphics[width=0.5\linewidth]{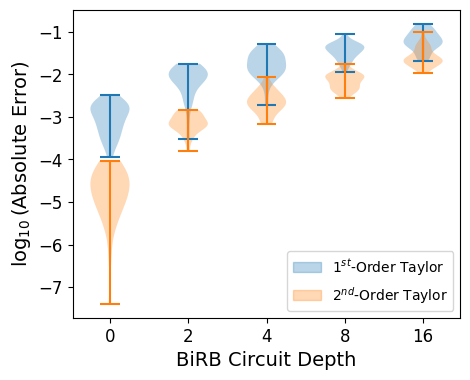}
    \caption{Absolute error between approximately and exactly computed BiRB circuit energies, $|\langle s_c \rangle_{\text{approx}} -  \langle s_c \rangle_{\text{exact}}|$. Here we're comparing the accuracy of approximations using the first-order BCH approximation and both first-and second-order Taylor approximations. As observed in Fig. \ref{fig:birb_energy_exact_approx_combined}, while the second-order Taylor approximation generally outperforms the first-order approximation across the board, this is particularly impactful at higher circuit depths.}
    \label{fig:birb_circuit_abs_error}
\end{figure}

To quantify the quality of the simulated results more quantitatively we've computed the absolute error between the exact and approximately computed BiRB circuit energies, $|\langle s_c \rangle_{\text{approx}} -  \langle s_c \rangle_{\text{exact}}|$. The absolute errors are plotted in Fig. \ref{fig:birb_circuit_abs_error} as a function of circuit depth, validating quantitatively the improvement in simulation accuracy seen in Fig. \ref{fig:birb_energy_exact_approx_combined} from increasing the order of our approximations. Randomized benchmarking is one of the most widely-deployed QCVV protocols experimentally, so the ability to accurately and efficiently simulate the behavior of RB circuits under physically-realistic noise environments as we've demonstrated here will likely prove incredibly valuable in refining the diagnostic utility of RB.

\newpage
\section{Formulae for efficiently manipulating elementary error generators}\label{sm:eegs}
Here we provide the various formulae that we use for efficiently manipulating the basis of EEGs, as well as those derivations of these formulae referred to in the main text.

\subsection{Conjugation of an elementary error generator by a Clifford superoperator}\label{sec:conjugations}

In this section we show how any elementary error generator is transformed under conjugation by a Clifford superoperator. Let $\mathbb{P}$ denote the set of all $4^n$ $n$-qubit Pauli operators without signs (e.g., for $n=1$, $\mathbb{P}=\{I, X, Y, Z\}$). Let $U$ denote some Clifford unitary, which is all unitaries such that $U P U^{\dagger} = + Q$ or $U P U^{\dagger} = - Q$ for some $Q \in \mathbb{P}$. As introduced in the main text, it will therefore be useful to define
$P_U \in \mathbb{P}$ and $s_{U,P} \in \{-1,+1\}$ to satisfy
\begin{equation}
 U^{\dagger} P U = s_{U,P} P_U.
\end{equation}
As throughout, let $\mathcal{U}$ be $U$'s superoperator representation, i.e., $\mathcal{U}[\rho] = U \rho U^{\dagger}$. 
We are now ready to derive $\mathcal{U}^{\dagger} \eg{G} \mathcal{U}$ for an arbitrary elementary error generator $\eg{G}$. We begin with the Hamiltonian error generators:
\begin{align}
     \mathcal{U}^{\dagger}  \eg{H}_P \mathcal{U}\left[\rho\right] & =-i U^{\dagger}\left[P,U\rho U^{\dagger}\right] U \\
     &= -i ( U^{\dagger}PU\rho U^{\dagger} U -  U^{\dagger}  U\rho U^{\dagger} P U)\\
     & = -i s_{U,P}( P_U\rho  - \rho P_U)\\
     & = -i s_{U,P}[P_U, \rho]\\
     & = s_{U,P} H_{P_U}[\rho].
 \end{align}
Next, we consider the stochastic error generators:
  \begin{align}
     \mathcal{U}^{\dagger}  \eg{S}_P \mathcal{U}\left[\rho\right] &= U^{\dagger}(P U \rho U^{\dagger} P- U\rho U^{\dagger}) U\\
     &=  (U^{\dagger}P U) \rho (U^{\dagger} P U) - \rho\\
     &=  s_{U,P}^2 P_U \rho P_U - \rho\\
     &=  \eg{S}_{P_U}[\rho].
 \end{align}
To consider the correlated and active error generators, we will need the following equation:
 \begin{equation}
   U^{\dagger} \left\{P,Q\right\} U = \{U^{\dagger}PU,U^{\dagger}Q U \} = s_{U,P}s_{U,Q} \{P_U,Q_U\},
   \end{equation}
   as well as the equivalent equation for the commutator:
    \begin{equation}
   U^{\dagger} \left[P,Q\right] U = [U^{\dagger}PU,U^{\dagger}QU ] = s_{U,P}s_{U,Q}[P_U,Q_U].
   \end{equation}
For the correlated error generators, we have that:
 \begin{align}
 \mathcal{U}^{\dagger} \eg{C}_{P,Q} \mathcal{U} \left[\rho\right]&= U^{\dagger} \left( P U\rho U^{\dagger} Q  +   QU\rho U^{\dagger} P  -  \frac{1}{2}\left\{\left\{P,Q\right\}, U\rho U^{\dagger}\right\} \right)U ,\\
 &= U^{\dagger} P U\rho U^{\dagger} Q U +  U^{\dagger} QU\rho U^{\dagger} P U - \frac{1}{2}\left\{  U^{\dagger} \left\{P,Q\right\} U, \rho  \right\}   \\
    &= s_{U,P}s_{U,Q}\left( P_U \rho Q_U +  Q_U \rho P_U - \frac{1}{2}\left\{  \{P_U,Q_U\}, \rho \right\}\right) \\
     &= s_{U,P}s_{U,Q}\eg{C}_{P_U,Q_U}[\rho].
  \end{align}
 Finally, we consider the active error generators:
  \begin{align}
  \mathcal{U}^{\dagger}  \eg{A}_{P,Q}  \mathcal{U}\left[\rho\right]&= i U^{\dagger} \left(P U\rho U^{\dagger} Q-Q U \rho U^{\dagger} P+\frac{1}{2}\left\{\left[P,Q\right],U \rho U^{\dagger} \right\}\right)U\\ 
  &= i \left( U^{\dagger} P U\rho U^{\dagger} Q U - U^{\dagger}Q U \rho U^{\dagger} P  U +\frac{1}{2}\left\{U^{\dagger}\left[P,Q\right]U, \rho \right\} \right)\\
    &= i  s_{U,P}s_{U,Q} \left(P_U\rho Q_U - Q_U \rho P_U +\frac{1}{2}\left\{\left[P_U,Q_U\right], \rho \right\} \right)\\
      &= s_{U,P}s_{U,Q}\eg{A}_{P_U,Q_U}[\rho].
   \end{align}

\subsection{Commutation relations for pairs of elementary error generators}\label{sec:commutators}
\label{SI:ComRel}
In this section we present the commutation relations between all pairs of EEGs, which are defined in the main text. Note that the right hand sides of these equations use an expanded definition of the EEG where we let the indexing Pauli operators be signed. We note the following identities between EEGs indexed with signed Pauli operators and the standard definitions of the EEGs with unsigned Pauli operators:
\begin{align}
\eg{H}_{w_P P}&=w_P\eg{H}_P\\
\eg{S}_{w_P P}&=w_P^2\eg{S}_P\\
\eg{C}_{w_P P,w_Q Q}&=w_P w_Q\eg{C}_{P,Q}\\
\eg{A}_{w_P P,w_Q Q}&=w_P w_Q\eg{A}_{P,Q},
\end{align}
where $w_P, w_Q \in \{\pm 1\}$ and $P,Q \in \mathbb{P}^*$. We have:
\begin{align*}
\left[\eg{H}_P,\eg{H}_Q\right] &= -i\eg{H}_{[P,Q]} \\
\left[\eg{H}_P,\eg{S}_Q\right] &= i \eg{C}_{Q,[Q,P]} \\
 \left[\eg{H}_Q,\eg{C}_{A,B}\right] &= i (\eg{C}_{[A,Q],B}+  \eg{C}_{[Q,B],A}) \\
\left[\eg{H}_{P},\eg{A}_{A,B}\right] &= -i(\eg{A}_{[P,A],B}+\eg{A}_{A,[P,B]}) \\
\left[\eg{S}_P,\eg{S}_Q\right] &= 0 \\
\left[\eg{S}_P,\eg{C}_{A,B}\right] & = -i\eg{A}_{AP,BP}-i\eg{A}_{PB,AP}+\frac{-i}{2}\left(\eg{A}_{\{A,B\}P,P}+\eg{A}_{P,P\{A,B\}}\right) \\
[\eg{S}_P,\eg{A}_{A,B}] &= i\left(\eg{C}_{PA,BP}-\eg{C}_{PB,AP}\right)-\frac{i}{2}\eg{A}_{P,[P,[A,B]]} \\ 
[\eg{C}_{A,B},\eg{C}_{P,Q}] &=-i\left(\eg{A}_{AP,QB}+\eg{A}_{AQ,PB}+\eg{A}_{BP,QA}+\eg{A}_{BQ,PA}\right)-\frac{i}{2}\left(\eg{A}_{[P,\{A,B\}],Q}+\eg{A}_{[Q,\{A,B\}],P}+\eg{A}_{[\{P,Q\},A],B}+\eg{A}_{[\{P,Q\},B],A}\right)+\frac{i}{4}\eg{H}_{[\{A,B\}\{P,Q\}]} \\
\left[\eg{C}_{A,B},\eg{A}_{P,Q}\right]&=i\left(\eg{C}_{AP,QB}-\eg{C}_{AQ,PB}+\eg{C}_{BP,QA}-\eg{C}_{PA,BQ}\right)
+\frac{1}{2}\left(\eg{A}_{[A,[P,Q],B}+\eg{A}_{[B,[P,Q]],A}+i\eg{C}_{[P,\{A,B\}],Q}-\eg{C}_{[Q,\{A,B\}],P}\right) -\frac{1}{4}\eg{H}_{[[P,Q],\{A,B\}]} \\
\left[\eg{A}_{A,B},\eg{A}_{P,Q}\right] &= -i\left(\eg{A}_{QB,AP}+\eg{A}_{PA,BQ}+\eg{A}_{BP,QA}+\eg{A}_{AQ,PB}\right) +\frac{1}{2}\left(\eg{C}_{[B,[P,Q]],A}-\eg{C}_{[]A,[P,Q]],B}+\eg{C}_{[P,[A,B]],Q}-\eg{C}_{[Q,[A,B]],P}\right)+\frac{i}{4}\eg{H}_{[[P,Q],[A,B]]}
\end{align*}

\subsection{Computational basis measurements on stabilizer states evolved by elementary error generators}\label{sec:alphas}
In this section we derive formulae for $\alpha(x, G, \psi)$, defined in Eq.~\eqref{eq:alpha}, and reproduced here:
\begin{equation}
\alpha(x, G, \psi) =  2^{\zeta(\psi)}   \textrm{Tr} (\ket{x}\bra{x} G[\ket{\psi}\bra{\psi}]).
\end{equation}
Here $\ket{\psi} = U_{\circuit}\ket{00\dots}$ is a stabilizer state, $G$ is one of the EEGs, and $x$ is an $n$ bit string. 
For any stabilizer state $\ket{\psi}$ we have that
\begin{equation}
    p_x =  \begin{cases} \frac{1}{2^{\zeta(\psi)}}  & \textrm{if $x \in \mathbb{Z}_\psi$}\\
    0  & \textrm{else}\\
    \end{cases}
\end{equation}
where $\mathbb{Z}_{\psi}$ is a $2^{\zeta(k)}$-element subset of all $n$-qubit strings $\mathbb{Z}$. It is efficient to compute whether $x$ is in $\mathbb{Z}_{\psi}$, using the stabilizer representation of $\ket{\psi}$, for any $x$ and $\psi$. For any two bit strings $x,y \in \mathbb{Z}_{\psi}$, let
\begin{equation}
 \Phi_{\psi,x}(P,Q) := 2^{\zeta(\psi)} \bra{x}P\ket{\psi}\bra{\psi}Q\ket{x}.
\end{equation}
Note that $\Phi_{\psi}(x,y)$ is invariant under multiplication of $\ket{\psi}$ by a global phase, and that $\Phi_{\psi}(x,y)$ can be efficiently computed for any $x$, $y$ and $\psi$. Because $\ket{\psi}$ is a stabilizer state it is an equal superposition of those computational basis states $\ket{x}$ states where $x \in \mathbb{Z}_{\psi}$ each multiplied by a phase of $\pm 1$ or $\pm i$, with an arbitrary global phase. This implies that 
\begin{equation}
 \Phi_{\psi}(P, Q) \in \{0, \pm i, \pm 1\}.
\end{equation}

Each $\alpha(x, G, \psi)$ can be expressed in terms of $\Phi_{\psi,x}(P,Q)$ as:
\begin{align}
  \alpha(x, \eg{S}_P, \psi)  & = \Phi_{\psi,x}(P,P) - \Phi_{\psi,x}(I,I) \label{alpha-S} \\ 
    \alpha(x, \eg{H}_P, \psi) &=  2 \Im \left[  \Phi_{\psi,x}(P, I) \right] \label{alpha-H} \\
     \alpha(x, \eg{C}_{P,Q}, \psi)  &= 2 \Re\left(\Phi_{\psi,x}(P,Q)\right)  - \Re\left(\Phi_{\psi,x}(PQ,I) + \Phi_{\psi,x}(QP,I)\right) \label{alpha-C} \\
          \alpha(x, \eg{A}_{P,Q}, \psi)  &=  2\Im\left( \Phi_{\psi,x}(Q,P) \right) + \Im\left( \Phi_{\psi,x}(QP,I) - \Phi_{\psi,x}(PQ,I)\right) \label{alpha-A}
\end{align}

\subsubsection{Derivations}
We now derive Eqs.~\eqref{alpha-S}-\eqref{alpha-A} for $\alpha(x, G, \psi)$. We begin with Eq.~\eqref{alpha-S}, where $G = \eg{S}_P$. In this case
\begin{align}
  \alpha(x, \eg{S}_P, \psi) = 2^{\zeta(\psi)} \textrm{Tr} (\ket{x}\bra{x}\eg{S}_P[\ket{\psi}\bra{\psi}])
& =  2^{\zeta(\psi)} \left(\bra{x}P\ket{\psi}\bra{\psi}P \ket{x} -  \braket{x}{\psi}\braket{\psi}{x} \right)\\
& = \Phi_{\psi,x}(P,P) - \Phi_{\psi,x}(I,I).
\end{align}

Next, we derive Eq.~\eqref{alpha-H}, which is a formula for $\alpha(x, \eg{H}_{P}, \psi)$. We have that:
\begin{align}
 \alpha(x, \eg{H}_P, \psi) &= 2^{\zeta(\psi)} \textrm{Tr} (\ket{x}\bra{x}\eg{H}_P[\ket{\psi}\bra{\psi}]) \\
& = 2^{\zeta(\psi)}i \left(\braket{x}{\psi}\bra{\psi}P \ket{x} -  \bra{x} P\ket{\psi}\braket{\psi}{x}\right)\\
& = i \left( \Phi_{\psi,x}(I, P) - \Phi_{\psi}(P, I)\right)\\
& = -i \left(\Phi_{\psi}(P, I)-  \Phi^*_{\psi,x}(P, I)\right)\\
& = 2\Im \left(\Phi_{\psi, x}(P, I)\right).
\end{align}

Now we derive Eq.~\eqref{alpha-C}, which is a formula for $\alpha(x, \eg{C}_{P,Q}, \psi)$. $P$ and $Q$ must either commute or anti-commute (i.e., $QP=-PQ$, so that $\{Q,P\}=0$). First we consider the case in which $P$ and $Q$ anti-commute. In this case $\eg{C}_{P,Q}[\rho] = P\rho Q + Q \rho P$, and so
\begin{align}
 \alpha(x, \eg{C}_{P,Q}, \psi) 
 &= 2^{\zeta(\psi)} \textrm{Tr}(\ket{x}\bra{x}\eg{C}_{P,Q}[\ket{\psi}\bra{\psi}])\\
 &= 2^{\zeta(\psi)}\left( \bra{x}P \ket{\psi}\bra{\psi} Q \ket{x} + \bra{x}Q \ket{\psi}\bra{\psi} P \ket{x} \right) \\ 
 &= \Phi_{\psi,x}(P,Q) + \Phi_{\psi,x}(Q,P) \\ 
  &= \Phi_{\psi,x}(P,Q) + \Phi^*_{\psi,x}(P,Q) \\ 
  &=  2 \Re\left(\Phi_{\psi,x}(P,Q)\right). 
\end{align}
We now consider the case in which $P$ and $Q$ commute, in which case   $\eg{C}_{P,Q}[\rho] = P\rho Q + Q \rho P + PQ \rho + \rho PQ $, and so
\begin{align}
 \alpha(x, \eg{C}_{P,Q}, \psi) 
 &= 2^{\zeta(\psi)} \textrm{Tr}(\ket{x}\bra{x}\eg{C}_{P,Q}[\ket{\psi}\bra{\psi}])\\
 &= 2^{\zeta(\psi)}\left( \bra{x}P \ket{\psi}\bra{\psi} Q \ket{x} + \bra{x}Q \ket{\psi}\bra{\psi} P \ket{x} -  \bra{x}PQ \ket{\psi}\braket{\psi}{x}   -  \braket{x}{\psi}\bra{\psi}QP\ket{x}  \right) \\ 
  &= 2^{\zeta(\psi)}\left( \Phi_{\psi,x}(P,Q) + \Phi^*_{\psi,x}(P,Q) -  \Phi_{\psi,x}(PQ,I)   -  \Phi^*_{\psi,x}(PQ,I)  \right) \\ 
  & = 2 \Re\left(\Phi_{\psi,x}(P,Q)\right)  -   2 \Re\left(\Phi_{\psi,x}(PQ,I)\right).
\end{align}
To encompass both the cases when $P$ and $Q$ commute and when they anti-commute, we can write
\begin{align}
 \alpha(x, \eg{C}_{P,Q}, \psi)  = 2 \Re\left(\Phi_{\psi,x}(P,Q)\right)  - \Re\left(\Phi_{\psi,x}(PQ,I) + \Phi_{\psi,x}(QP,I)\right),
\end{align}
which is the expression given in Eq.~\eqref{alpha-C}.

Finally, we derive Eq.~\eqref{alpha-A}, which is a formula for $\alpha(x, \eg{A}_{P,Q}, \psi)$. $P$ and $Q$ must either commute or anti-commute, and first we consider the case in which $P$ and $Q$ commute. In this case $\eg{A}_{P,Q}[\rho] = i(P\rho Q - Q \rho P)$, and so
\begin{align}
 \alpha(x, \eg{A}_{P,Q}, \psi) 
 &= 2^{\zeta(\psi)} \textrm{Tr}(\ket{x}\bra{x}\eg{A}_{P,Q}[\ket{\psi}\bra{\psi}])\\
 &= 2^{\zeta(\psi)}i\left( \bra{x}P \ket{\psi}\bra{\psi} Q \ket{x} - \bra{x}Q \ket{\psi}\bra{\psi} P \ket{x} \right) \\ 
 &=i\left( \Phi_{\psi,x}(P,Q) - \Phi_{\psi,x}(Q,P) \right) \\ 
 &=-i\left( \Phi_{\psi,x}(Q,P) - \Phi_{\psi,x}^*(Q,P) \right)
  \\ 
 &=2\Im\left( \Phi_{\psi,x}(Q,P)\right)
\end{align}
Now we consider the case in which $P$ and $Q$ do not commute, so they anti-commute. In this case $\eg{A}_{P,Q}[\rho] = i(P\rho Q - Q \rho P + PQ\rho - \rho QP)$, and so
\begin{align}
 \alpha(x, \eg{A}_{P,Q}, \psi) 
 &= 2^{\zeta(\psi)} \textrm{Tr}(\ket{x}\bra{x}\eg{A}_{P,Q}[\ket{\psi}\bra{\psi}])\\
 &= 2^{\zeta(\psi)}i\left( \bra{x}P \ket{\psi}\bra{\psi} Q \ket{x} - \bra{x}Q \ket{\psi}\bra{\psi} P \ket{x} +  \bra{x}PQ \ket{\psi}\braket{\psi}{x} - \braket{x}{\psi}\bra{\psi} QP \ket{x}  \right) \\ 
 &=-i\left( \Phi_{\psi,x}(Q,P) - \Phi_{\psi,x}^*(Q,P) + \Phi_{\psi,x}(QP,I) - \Phi^*_{\psi,x}(QP,I) \right) \\ 
 &=2\Im\left( \Phi_{\psi,x}(Q,P) + \Phi_{\psi,x}(QP,I)\right)
\end{align}
To encompass both the cases when $P$ and $Q$ commute and when they anti-commute, we can write
\begin{align}
 \alpha(x, \eg{A}_{P,Q}, \psi)  = 2\Im\left( \Phi_{\psi,x}(Q,P) \right) + \Im\left( \Phi_{\psi,x}(QP,I) - \Phi_{\psi,x}(PQ,I)\right)
\end{align}
which is the expression given in Eq.~\eqref{alpha-A}.

\subsection{Pauli expectation values for stabilizer states evolved by elementary error generators}
\label{sec:betas}

In this section we derive expressions for $\beta(\psi, G, P) = \Tr(PG[\ketbra{\psi}])$. First we consider the case of an arbitrary stochastic EEG $G = S_Q$ with $Q \in \mathbb{P}^*$. We have that:

\begin{align}
    \Tr(PG[\ketbra{\psi}]) & = \Tr(PS_Q[\ketbra{\psi}]) \\
    & = \Tr(PQ \ketbra{\psi} Q) - \Tr(P \ketbra{\psi}) \\
    & = \bra{\psi} QPQ \ket{\psi} - \bra{\psi} P \ket{\psi}
\end{align}
Therefore, $\Tr(PG[\ketbra{\psi}])=0$ if $[P,Q]=0$. Otherwise, $\Tr(PG[\ketbra{\psi}]) = -2\bra{\psi} P \ket{\psi}$. If $P\ket{\psi} = \ket{\psi}$ (i.e, $P$ stabilizes $\ket{\psi}$) then  $\Tr(PG[\ketbra{\psi}]) = -2$. If $P\ket{\psi} = -\ket{\psi}$ (i.e, $P$ anti-stabilizes $\ket{\psi}$) then $\Tr(PG[\ketbra{\psi}]) = 2$. If $P$ does not stabilize or anti-stabilize $\ket{\psi}$ then $\bra{\psi} P\ket{\psi} = 0$, so $\Tr(PG[\ketbra{\psi}]) = 0$. Therefore, in summary:
\begin{equation}
   \beta(\psi, S_Q, P) =  \begin{cases} \pm 2 & \textrm{if } [P,Q] \neq 0 \textrm{ and  } P\ket{\psi} = \mp \ket{\psi}\\
    0  & \textrm{else}\\
    \end{cases}
\end{equation}

Next we consider the case of an arbitrary Hamiltonian EEG: $G = H_Q$ with $Q \in \mathbb{P}^*$. We have that

\begin{align}
    \Tr(PG[\ketbra{\psi}]) & = \Tr(PH_Q[\ketbra{\psi}]) \\
    & = i\Tr(PQ \ketbra{\psi}) - i\Tr(P \ketbra{\psi} Q) \\
    & = i\bra{\psi} [P,Q] \ket{\psi} 
\end{align}
Therefore,  $\Tr(PG[\ketbra{\psi}])=0$ if $[P,Q]=0$. Otherwise $\{P,Q\}=0$, implying $i[P,Q]=2iPQ = 2Q'$ where $Q'$ is a signed Pauli operator (a Pauli operator multiplied by either $+1$ or $-1$), and so $\Tr(PG[\ketbra{\psi}])= 2\bra{\psi} Q' \ket{\psi} $. If $Q'\ket{\psi} = \pm \ket{\psi}$ (i.e, $Q'$ stabilizes or anti-stabilizes $\ket{\psi}$) then $\Tr(PG[\ketbra{\psi}]) = \pm 2$. If $Q'$ does not stabilize or anti-stabilize $\ket{\psi}$ then $\bra{\psi} Q'\ket{\psi} = 0$, so $\Tr(PG[\ketbra{\psi}]) = 0$. Therefore, in summary

\begin{equation}
   \beta(\psi, H_Q, P) =  \begin{cases} \pm 2 & \textrm{if } [P,Q] \neq 0 \textrm{ and  } iPQ\ket{\psi} = \pm \ket{\psi}\\
    0  & \textrm{else}\\
    \end{cases}
\end{equation}

Next, we consider the case of an arbitrary correlation EEG: $G = C_{Q_1, Q_2}$ for $Q_1 \neq Q_2$. We have that 
\begin{align}
    \Tr(PG[\ketbra{\psi}]) & = \Tr(PC_{Q_1, Q_2}[\ketbra{\psi}]) \\
    & = \Tr(PQ_1 \ketbra{\psi} Q_2) + \Tr(PQ_2 \ketbra{\psi} Q_1) + \Tr(P(\{Q_1Q_2, \ketbra{\psi}\}+\{Q_2Q_1, \ketbra{\psi}\})) \\
    & = \Tr(PQ_1 \ketbra{\psi} Q_2) + \Tr(PQ_2 \ketbra{\psi} Q_1) + \frac{1}{2}\Tr((PQ_1Q_2 + PQ_2Q_1 + Q_1Q_2P + Q_2Q_1P)\ketbra{\psi}) \\
    & = i\bra{\psi} Q_2PQ_1 \ket{\psi} - i\bra{\psi} Q_1PQ_2 \ket{\psi} + \frac{i}{2} \bra{\psi}\{P,\{Q_1, Q_2\}\} \ket{\psi})
\end{align}

If $[Q_1, Q_2]\neq 0$, then
\begin{align}
    \Tr(PC_{Q_1, Q_2}[\ketbra{\psi}]) = \bra{\psi} Q_2PQ_1 \ket{\psi} + \bra{\psi} Q_1PQ_2 \ket{\psi}.
\end{align}
Finally, we consider the case of an arbitrary active EEG: $G = C_{Q_1, Q_2}$ for $Q_1 \neq Q_2$. We have that 
\begin{align}
    \Tr(PG[\ketbra{\psi}]) & = \Tr(PA_{Q_1, Q_2}[\ketbra{\psi}]) \\
    & = i\Tr(PQ_1 \ketbra{\psi} Q_2)- i \Tr(PQ_2 \ketbra{\psi} Q_1) + \frac{i}{2}\Tr(P(\{Q_1Q_2, \ketbra{\psi}\}-\{Q_2Q_1, \ketbra{\psi}\})) \\
    & = i\Tr(PQ_1 \ketbra{\psi} Q_2) -i \Tr(PQ_2 \ketbra{\psi} Q_1) + \frac{i}{2}\Tr((PQ_1Q_2 + Q_1Q_2P - PQ_2Q_1 - Q_2Q_1P)\ketbra{\psi}) \\
    & = \bra{\psi} Q_2PQ_1 \ket{\psi} + \bra{\psi} Q_1PQ_2 \ket{\psi} + \frac{1}{2} \bra{\psi}\{P,[Q_1, Q_2]\} \ket{\psi})
\end{align}

\end{document}